\documentclass[useAMS,usenatbib,usegraphicx]{mn2e}

\usepackage{color, aas_macros}
\def\eref#1{equation (\ref{eq:#1})}
\def\eeref#1{(\ref{eq:#1})}
\def\fref#1{Fig. \ref{fig:#1}}
\def\sref#1{section \ref{sec:#1}}

\def\bcdot{\bmath{\cdot}}
\def\change#1{{#1}}

\title{Three-dimensional stability of magnetically confined mountains
  on accreting neutron stars}

\author[M. Vigelius et al.]{M.~Vigelius $^1$\thanks{E-mail: mvigeliu@physics.unimelb.edu.au} and A.~Melatos $^1$ \\
 $^1$ School of Physics, University of Melbourne, Parkville, VIC
 3010, Australia}

\begin{document}

\date{Submitted to MNRAS}

\maketitle

\begin{abstract}
  We examine the hydromagnetic stability of magnetically confined mountains,
  which arise when material accumulates at the magnetic poles of an
  accreting neutron star. We extend a previous
  axisymmetric stability analysis by performing three-dimensional
  simulations using the ideal-magnetohydrodynamic (ideal-MHD) code
  \textsc{zeus-mp}, investigating the role played by boundary
  conditions, accreted mass, stellar curvature, and (briefly) toroidal
  magnetic field strength. We find that axisymmetric equilibria are
  susceptible to the   undular sub-mode of the Parker instability but are
  not disrupted. The line-tying boundary condition at the stellar
  surface is crucial in stabilizing the mountain. The
  nonlinear three-dimensional saturation state of the instability is
  characterized by a small degree of  nonaxisymmetry ($\la 0.1$ per 
  cent) and a mass  ellipticity of $\epsilon \sim 10^{-5}$ for an
  accreted mass of $M_a = 10^{-5} M_\odot$. Hence there is
  a good prospect of detecting gravitational waves from accreting
  millisecond pulsars with long-baseline interferometers such as
  Advanced LIGO. We also investigate the ideal-MHD spectrum of the
  system, finding that long-wavelength poloidal
  modes are suppressed in favour of toroidal modes in the
  nonaxisymmetric saturation state.
\end{abstract}

\bibliographystyle{mn2e}


\section{Introduction}
There exists strong observational evidence that the
magnetic dipole moment of accreting neutron stars in X-ray
binaries, $\mu$, decreases with accreted mass, $M_a$
\citep{Taam86, vanDenHeuvel95}, although \citet{Wijers97} noted that
extra variables may enter this relation. Numerous mechanisms for field
reduction have been proposed, such as 
accelerated Ohmic decay \citep{Konar97,Urpin97}, vortex-fluxoid
interactions in the superconducting core \citep{Muslimov85,
  Srinivasan90}, and magnetic screening or burial
\citep{Bisnovatyi74,Romani90,Zhang98,Payne04,Lovelace2005}. The reader is
referred to \citet{Melatos01} for a comparative review.

Magnetic burial occurs when accreted plasma flowing inside the
Alfv\'{e}n radius is chanelled onto the
magnetic poles of the neutron star. The hydrostatic pressure at the
base of the accreted column overcomes
the magnetic tension of the magnetic field lines and spreads
equatorwards, thereby distorting the frozen-in magnetic flux
\citep{Melatos01}. \citet{Payne04}, hereafter PM04, computed self-consistently the
\emph{unique} quasistatic sequence of ideal-MHD equilibria that
describes how burial proceeds as a function
of $M_a$, \emph{while respecting the flux freezing constraint} of
ideal MHD. They found that the magnetic field is compressed into an equatorial belt,
which confines the accreted mountain at the poles. A key result is
that $\mu$ is reduced significantly once $M_a$ exceeds $\sim 10^{-5}
M_\odot$, five orders of magnitude above previous 
estimates \citep{Brown98, Litwin01}.

Generally speaking, one expects highly distorted hydromagnetic equilibria
like those in PM04 to be disrupted on the Alfv\'{e}n time-scale by a
plethora of MHD instabilities. Surprisingly, however,
\citet{Payne06b} (hereafter PM07) found the equilibria to be stable to
axisymmetric ideal-MHD modes. When kicked, the mountain performs radial
and lateral oscillations corresponding to global Alfv\'{e}n and
compressional modes, but it remains intact.

An axisymmetric analysis, however, excludes instabilities
involving toroidal modes and is therefore
incomplete. In general, three-dimensional effects alter MHD stability,
quantitatively and qualitatively. For example, \citet{Matsumoto92} found
the growth rate of the three-dimensional Parker instability to be
higher than that of its two-dimensional counterpart.
\citet{Masada06} proved that newly born neutron stars containing a
toroidal field are stable to the axisymmetric magneto-rotational
instability yet unstable to its nonaxisymmetric
counterpart. Differences between the two- and three-dimensional
stability of MHD equilibria are also observed in a variety of solar
contexts \citep{Priest84} and in tokamaks \citep{Lifschitz89,
  Goedbloed04}.

The central aim of this paper is to perform
fully three-dimensional, ideal-MHD simulations to assess the stability
of magnetic mountains, generalizing PM07. Importantly, we compute not
just the linear growth rate but also the \emph{nonlinear saturation state} of
any unstable modes. The latter property is what matters over the long
accretion time-scale when evaluating magnetic burial as the cause of
the observed reduction in $\mu$ \citep{Payne05},  the persistence of
millisecond oscillations in type-I X-ray bursts \citep{Payne07c}, and
gravitational radiation from magnetic mountains
\citep{Melatos05, Payne06a}. 

The structure of the paper is as
follows. We introduce our numerical setup in \sref{model} and validate it against previous
axisymmetric results in \sref{axisymmetric}, characterizing the
controlling influence of the boundary conditions for the first
time. In \sref{stab3d}, we present three-dimensional simulations,
which display growth of unstable toroidal modes. The instability is
classified according to its dispersive properties and energetics, and
the nonlinear saturation state is computed as a function of $M_a$. We
compute the spectrum of global MHD oscillations in
\sref{oscillations}. Resistive effects are postponed to a future paper.

\section{Numerical model}
\label{sec:model}

\change{
The accretion problem contains two fundamentally different
time-scales: the long accretion time ($\sim 10^8$ yr) and the
short Alfven time ($\sim 10^{-3}$ s). The wide discrepancy prevents
us from treating the accretion problem dynamically,
i.e. in a full MHD simulation, where mass is added through
the outer boundary onto an initially dipolar field. Instead, for a
given value of $M_a$, we compute the magnetohydrostatic equilibria,
using the Grad-Shafranov solver developed by PM04, then load it
into the ideal-MHD solver \textsc{zeus-mp} \citep{Hayes06} to test its
hydromagnetic stability on the Alfv\'{e}n time-scale $\tau_\mathrm{A}$.
We find below that all quantities reach their saturation values after $\sim
10 \tau_A \sim 10^{-2}$ s at a particular value of $M_a$ (e.g. $10^{-4}
M_\odot$). As $M_a$ changes slowly, over $\sim 10^8$ yr, the saturation
values adjust in a quasistatic way on the Alfv\'{e}n time-scale. In
practice, to study a different value of $M_a$ numerically, we
recalculate the Grad-Shafranov eqilibrium and load the new equilibrium
into \textsc{zeus-mp}.
}

\subsection{Magnetic mountain equilibria}
\label{sec:model:mountains}
Analytic and numerical recipes for calculating
self-consistent ideal-MHD equilibria for magnetic mountains are
set out in PM04. Here, we briefly
restate the main points for the convenience of the reader.

An axisymmetric equilibrium is generated by a scalar flux function $\psi(r,
\theta)$, such that the magnetic field
\begin{equation}
  \mathbf{B}=\frac{\nabla \psi}{r \sin \theta} \times \hat{\mathbf{e}}_\phi
\end{equation}
automatically satisfies $\nabla \mathbf{\bcdot} \mathbf{B}=0$.
We employ the usual spherical coordinates $(r, \theta, \phi)$,
where $\theta=0$ corresponds to the symmetry axis of the magnetic
field before accretion. In the static limit, the mass conservation and
MHD induction equations are identically satisfied, while
the component of the momentum equation transverse to $\mathbf{B}$
reduces to a second order, nonlinear, elliptic partial differential
equation for $\psi$, the Grad-Shafranov (GS) equation (PM04):
\begin{equation}
  \label{eq:mountains:gs_master}
  \Delta^2 \psi = -F'(\psi) \exp[-(\varphi-\varphi_0)/c_s^2],
\end{equation}
with
\begin{equation}
  \Delta^2 = \frac{1}{\mu_0 r^2 \sin^2 \theta} \left[
    \frac{\partial^2}{\partial r^2}+\frac{\sin \theta}{r^2}
    \frac{\partial}{\partial \theta} \left( \frac{1}{\sin \theta}
      \frac{\partial}{\partial \theta} \right) \right].
\end{equation}

Formally, $F(\psi)$ is an arbitrary function. However,
the ideal-MHD flux-freezing constraint, that matter cannot cross flux surfaces,
imposes an additional conservation law on the mass-flux ratio
$\mathrm{d}M/\mathrm{d}\psi$,
\begin{equation}
  \label{eq:mountains:gs_supplement}
  F(\psi)=\frac{c_s^2}{2 \pi} \frac{\mathrm{d}M}{\mathrm{d}\psi}
  \left\{\int \mathrm{d}s \, r \sin\theta |\nabla \psi|^{-1}
    \mathrm{e}^{-(\varphi-\varphi_0)/c_s^2} \right\}^{-1},
\end{equation}
which determines $F(\psi)$ uniquely when solved simultaneously with
\eeref{mountains:gs_master}. The integration in
\eeref{mountains:gs_supplement} is performed along the
field line $\psi=\mathrm{const}$, $\phi=\mathrm{const}$.
The solution is insensitive to the exact form of
$\mathrm{d}M/\mathrm{d}\psi$; we thus distribute the 
accreted mass $M_a$ uniformly over $0 \leq \psi \leq \psi_a$, where
$\psi_a$ is the flux enclosed within the polar cap, and take $\psi$ to
be dipolar, initially, with hemispheric flux $\psi_\ast$.

In writing \eeref{mountains:gs_master} and \eeref{mountains:gravity}, we approximate the
gravitational field as uniform over the height of the mountain, and
hence write the gravitational potential $\varphi$ as
\begin{equation}
  \label{eq:mountains:gravity}
  \varphi = G M_\ast r/R_\ast^2,
\end{equation}
with $\varphi_0=G M_\ast/R_\ast$, where $M_\ast$ and $R_\ast$ denote the stellar mass and radius
respectively. We also assume an isothermal equation of state,
$\rho=c_s^2 p$, where $c_s$ denotes the sound speed.

By working in the ideal-MHD limit, we neglect elastic stresses
\citep{Melatos01, Haskell06, Owen06}, the Hall drift \citep{Geppert02,
  Cumming04, Pons07}, and Ohmic diffusion \change{\citep{Romani90, Geppert94}}. In particular, Ohmic diffusion
causes the mass quadrupole moment of the magnetic mountain (and $\mu$)
to saturate above a certain value of $M_a$ and may also affect the
stability to resistive MHD (e.g. ballooning) modes.\footnote{J. Arons, private
  communication.} We defer investigating these resistive effects
to a forthcoming paper [Vigelius \& Melatos (in preparation)].

We solve \eeref{mountains:gs_master} and
\eeref{mountains:gs_supplement} simultaneously using the relaxation
algorithm described in PM04, subject to the boundary conditions
$\psi(R_\ast, \theta)=\psi_\ast \sin^2 \theta$ (line tying at the surface),
$\partial \psi/\partial r(R_m, \theta)=0$ at the outer boundary $r=R_m$,
$\psi(r, 0)=0$, and $\partial \psi/\partial \theta(r,
\pi/2)=0$ (north-south symmetry).

\subsection{Evolution in ZEUS-MP}
\label{sec:model:zeus}
In this paper, we explore numerically how the axisymmetric GS
equilibria evolve when subjected to a variety of initial and
boundary conditions in three dimensions. To achieve this, we
employ the parallelized, general purpose, time-dependent, ideal-MHD
solver \textsc{zeus-mp} \citep{Hayes06}. \textsc{zeus-mp} integrates
the equations of ideal MHD, discretized on a fixed staggered grid. The
hydrodynamic part is based on a finite-difference advection
scheme accurate to second order in time and space. The magnetic tension
force and the induction equation are solved via the method of
characteristics and constrained transport (MOCCT)
\citep{Hawley95}, whose numerical implementation in \textsc{zeus-mp} is
described in detail by \citet{Hayes06}.

We initialize \textsc{zeus-mp} with an equilibrium computed by the
\textsc{GS} code and described by $\mathbf{B}(r, \theta)$ and $\rho(r, \theta)$, 
rotated about the $z$ axis to generate cylindrical
symmetry. We introduce initial
perturbations by taking advantage of the numerical
noise produced by the transition between grids in the GS code and
\textsc{zeus-mp}.

We adopt dimensionless variables in \textsc{zeus-mp} satisfying
$\mu_0=G=c_s=h_0=1$, where $h_0=c_s^2 R_\ast^2/G M_\ast$ denotes the
hydrostatic scale height. The basic units of mass,
magnetic field, and time are then $M_0=h_0 c_s^2/G$, \change{$B_0=[c_s^4/(G
h_0^2)]^{1/2}$}, and $\tau_0=h_0/c_s$. The grid and
boundary conditions are specified in appendix \ref{sec:app:zeusvars}.

\subsection{Curvature rescaling}
\label{sec:model:units}

In general, the characteristic length-scale for radial gradients
($h_0$) is much smaller than the length-scale for
latitudinal gradients $R_\ast$, creating numerical difficulties. However, in
the small-$M_a$ limit, it can be shown analytically (PM04, PM07) that
the structure of the magnetic mountain depends on
$R_\ast$ and $M_\ast$ through the combination $h_0 \propto
R_\ast^2/M_\ast$, not separately. We therefore artificially reduce $R_\ast$ and $M_\ast$, while
keeping $h_0$ fixed, to render the problem tractable computationally. It is
important to bear in mind that invariance of the equilibrium
structure under this curvature rescaling does not imply invariance
of the dynamical behaviour, nor is it necessarily applicable at large $M_a$.

A standard neutron star has $M_\ast=1.4 M_\odot$,
$R_\ast=10^6$ cm, $B_\ast=10^{12}$ G, and $c_s=10^8$ cm s$^{-1}$, giving
$h_0=53.82$ cm, $a=R_\ast/h_0=1.9 \times 10^4$, and $\tau_0=5.4\times 10^{-7}$ s. We
rescale the star to $M_\ast'=1.0\times10^{-5} M_\odot$ and
$R_\ast'=2.7 \times 10^3$ cm, reducing $a$ to 50 while keeping it
large. The base units for this rescaled star (see \sref{model:zeus})
are then $M_0=8.1\times10^{24}$ g, $\rho_0=5.2\times10^{19}$ g
cm$^{-3}$, \change{$B_0=7.2\times10^{17}$ G}, and $\tau_0=5.4\times 10^{-7}$ s. The
critical accreted mass above which the star's magnetic moment starts
to change, $M_c$, is defined by equation (30) of PM04:
\begin{equation}
  \frac{M_c}{M_\odot} = 6.2 \times 10^{-15} \left(\frac{a}{50}\right)^4
 \left( \frac{B_\ast}{10^{12} \mathrm{G}} \right)^2   
  \left( \frac{c_s}{10^8 \mathrm{cm\;s}^{-1}} \right)^{-4}.
\end{equation}

A characteristic time-scale for the MHD response of the mountain is
the Alfv\'{e}n pole-equator crossing time, $\tau_\mathrm{A}=\pi
R_\ast/(2 v_\mathrm{A})$, where $v_\mathrm{A}=(B^2/\mu_0 \rho)^{1/2}$
is the Alfv\'{e}n speed.  Clearly, $v_\mathrm{A}$ is
a function of position and time, so the definition of
$\tau_\mathrm{A}$ is somewhat arbitrary. Typically, at the
equator, we find $B \sim 10^{-6} B_0$ and $\rho \sim 10^{-11} \rho_0$, empirically
implying $\tau_\mathrm{A} \approx 250 \tau_0$.

\section{Axisymmetric stability and global oscillations}
\label{sec:axisymmetric}

\begin{table}
  \centering
  \caption{Simulation parameters. $M_a$ is the accreted mass, in units
    of the characteristic mass $M_c$ (\sref{model:units}), and $a=R_\ast/h_0$
  measures the curvature of the rescaled star
  (\sref{model:units}). The conditions at the outer boundary ($r=R_m$)
  are either \texttt{outflow} (zero gradient in all field variables)
  or \texttt{inflow} (pinned magnetic field); cf. also appendix \ref{sec:app:zeusvars}.}
  \begin{tabular}{@{}ccccp{0.1\textwidth}}
    \hline
    Model & $M_a/M_c$ & $a$ & Axisymmetry & Boundary \\
    \hline
    A & 1.0 & 50 & yes & outflow \\
    B & 1.0 & 50 & yes & inflow \\
    \hline
    D & 1.0 & 50 & no & outflow \\
    E & 1.0 & 50 & no & inflow \\
    \hline
    F & 0.6 & 50 & no & outflow \\
    G & 1.4 & 50 & no & outflow \\
    \hline
    J & 1.0 & 75 & no & outflow \\
    K & 1.0 & 100 & no & outflow \\
    \hline
\end{tabular}
  \label{tab:models}
\end{table}

PM07 demonstrated the axisymmetric stability of magnetic
mountains using the serial ideal-MHD solver \textsc{zeus-3d}. Here, we
start by repeating these axisymmetric simulations in the parallel
solver \textsc{zeus-mp}, in order to
verify the mountain implementation in \textsc{zeus-3d} and
\textsc{zeus-mp}, generate an axisymmetric reference model, and
understand the effect of the boundary conditions, which were not
investigated fully in previous work. The simulation parameters are
detailed in Table \ref{tab:models} (models A and B).

\begin{figure*}
  \includegraphics[width=168mm, keepaspectratio]{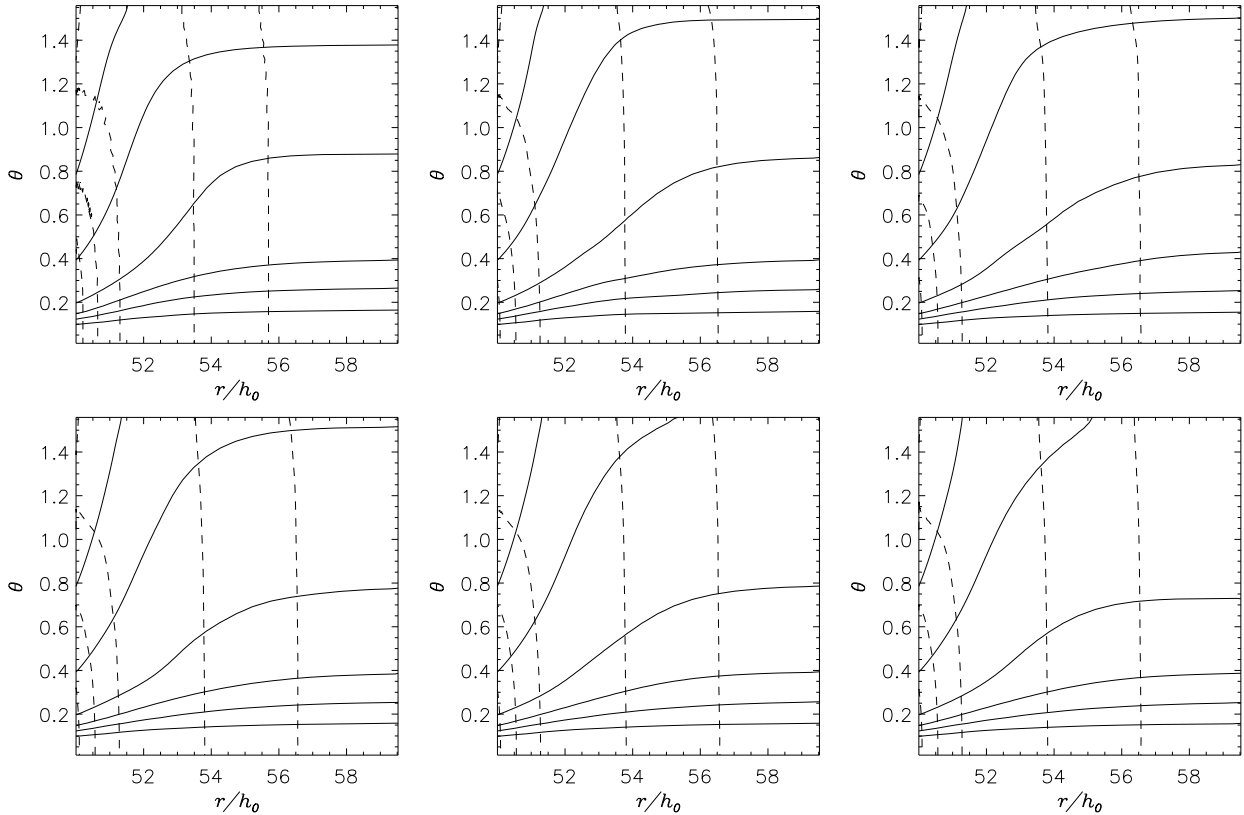}
  \caption{Meridional section of model A at $t/\tau_\mathrm{A}=0, 0.4,
    1, 1.8, 2.4, 3.6$ (top left to
    bottom right). Shown are density contours (dashed curves) with values
    $\log_{10}(\rho/\rho_0')=-13, -12, -11, -10.7, -10.5, -10.3$,
    and flux surfaces with footpoints at $r=R_\ast$,
    $\theta=0.10, 0.12,0.15, 0.20, 0.39, 0.79$  (solid curves). Lateral
    oscillations of the equatorial field lines are clearly
    visible. The \texttt{outflow} boundary condition at $r=R_m$
    makes the field lines flare towards the magnetic
    pole. \change{This is visible most clearly for
    the line whose footpoint lies at $\theta=0.39$}.}
  \label{fig:stab:moda2d}
\end{figure*}

\subsection{Reference model}
Model A, in which we set $M_a/M_c=1.0$ and the outer boundary condition
to \texttt{outflow},  serves as a reference case. \fref{stab:moda2d}
displays a time series of six $r$-$\theta$ sections for $0 \le
t/\tau_A \le 3.6$, showing density contours (dashed curves) and the
magnetic field lines projected into the plane $\phi=0$ 
(solid curves). The axisymmetric equilibrium (top-left panel) reveals
how the bulk matter is contained at the magnetic pole by the tension
of the distorted magnetic field.

The mountain in \fref{stab:moda2d} performs damped lateral oscillations without being
disrupted. The (unexpected) stability of this configuration is
due to two factors. First, the configuration is already the final, saturated
state of the nonlinear Parker instability, which is reached
quasistatically during slow accretion
\citep[PM04;][]{Mouschovias74}. Second, line-tying of the magnetic
field 
at $r=R_\ast$ significantly changes the structure of the MHD wave
spectrum in a way that enhances stability. \citet{Goedbloed94}
found that, in a homogenous plasma, a superposition of
Alfv\'{e}n- and magnetosonic waves is needed to satisfy the line-tying
boundary conditions. As a consequence, the basic, unmixed MHD modes are not
eigenfunctions of the linear force operator, and thus the spectrum is modified.

\begin{figure}
  \includegraphics[width=84mm, keepaspectratio]{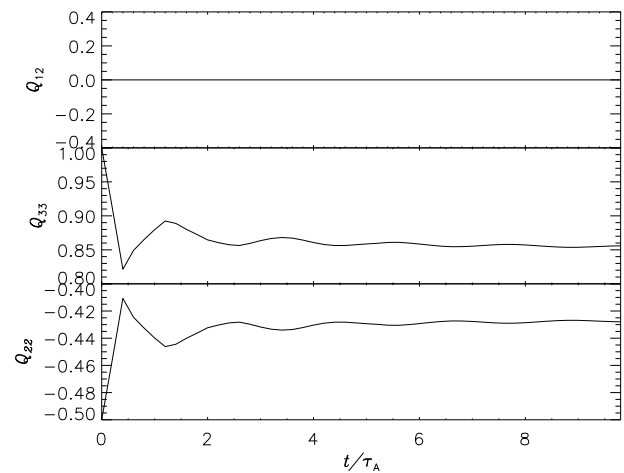}
  \caption{Mass quadrupole moments for model A, normalised to the maximum of $Q_{33}$ 
    ($1.3\times10^{25}$ g cm$^2$), as a function of time in units
    Alfv\'{e}n time. We find $Q_{22}=-Q_{33}/2$ and $Q_{12}=0$, as expected
    for an axisymmetric configuration. The global Alfv\'{e}n
    oscillation is damped by numerical viscosity.}
  \label{fig:stab:moda_quadru}
\end{figure}

The mass quadrupole moments,
\begin{equation}
  Q_{ij}=\int d^3 x' \, (3 x_i' x_j'-r'^2 \delta_{ij}) \rho(\mathbf{x'}),
\end{equation}
of the mountain in \fref{stab:moda2d} are plotted versus time in Fig.
\ref{fig:stab:moda_quadru}. We note first that $Q_{12}=0$ and
$Q_{22}=-Q_{33}/2$, as expected for an axisymmetric system. We can
compare Fig. \ref{fig:stab:moda_quadru} directly with the ellipticity
$\epsilon \propto Q_{33}$ computed by \citet{Payne06a}.\footnote{\citet{Payne06a}
  simulated a polar cap with $b=10$ as against $b=3$ in
  model A.}. These authors found two
dominant global modes, an Alfv\'{e}n and an acoustic mode, claimed to
be analogous to the fundamental modes of a gravitating, magnetized plasma slab. We
cannot resolve the compressional modes in \fref{stab:moda_quadru}, but
the latitudinal (Alfv\'{e}n) mode is clearly visible through the oscillations
in $Q_{22}$ and $Q_{33}$.

An oscillation cycle proceeds as follows. The first minimum of $Q_{33}$, and hence
$\epsilon$, at $t=0.4 \tau_\mathrm{A}$ in \fref{stab:moda_quadru}, corresponds to the top
right panel of \fref{stab:moda2d}. The mountain withdraws radially and
poleward. Polar field lines move closer to the magnetic pole, while
equatorial field lines are drawn towards the equator. At
$\tau=\tau_\mathrm{A}$, the mountain spreads and $Q_{33}$ reaches a
maximum in \fref{stab:moda_quadru}. The damping observed in
\fref{stab:moda_quadru} arises solely from numerical dissipation; neither
viscosity nor resistivity are included in our version of \textsc{zeus-mp}.

\subsection{Outer boundary}
The ``flaring up'' of magnetic field lines at the pole,
observed by PM07, is an artifact of the
\texttt{outflow} boundary condition at $r=R_m$. In order to check this,
we repeat the simulation of model A but switch to an \texttt{inflow}
boundary condition (model B). The density distribution is similar in
the two models, as is clear from \fref{stab:modb2d}. However, the
\texttt{inflow} BC artifically pins the magnetic field to the outer
boundary, introducing magnetic field discrepancies (mostly in the
outer layers, where $\rho$ is negligible).

\begin{figure}
  \includegraphics[width=84mm, keepaspectratio]{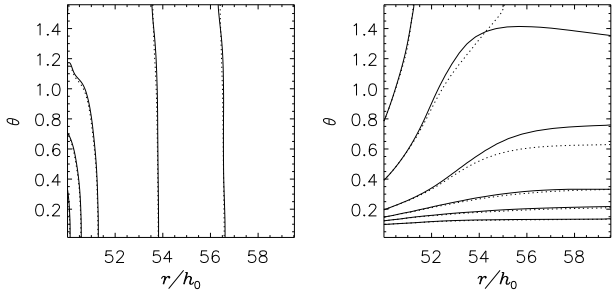}
  \caption{Meridional section of model B, showing density (left) and
    projected magnetic field (right) at $t/\tau_\mathrm{A}=3.6$ (solid curves). The
    field lines are pinned to the outer $r$ boundary by  the
    \texttt{inflow} boundary condition. For comparison, the density
    and magnetic field of model A are overplotted (dotted curves).}
  \label{fig:stab:modb2d}
\end{figure}

Alfv\'{e}n waves are similar to transverse waves on a string, where
the magnetic tension provides the restoring force. Models A and B are
therefore equivalent to a vibrating string with one end free and fixed
respectively. We expect the oscillation frequency of the
fundamental mode in model B to be twice that of model A. This is
indeed observed in the oscillations of the quadrupole moments,
displayed in \fref{stab:modb_quadru}: $Q_{22}$ and $Q_{33}$ in \fref{stab:modb_quadru} 
oscillate at $0.6$ times the period in \fref{stab:moda_quadru}.
A comprehensive analytic computation of the MHD spectrum, including
discrete and continuous components, will be attempted in a forthcoming paper.

\begin{figure}
  \includegraphics[width=84mm, keepaspectratio]{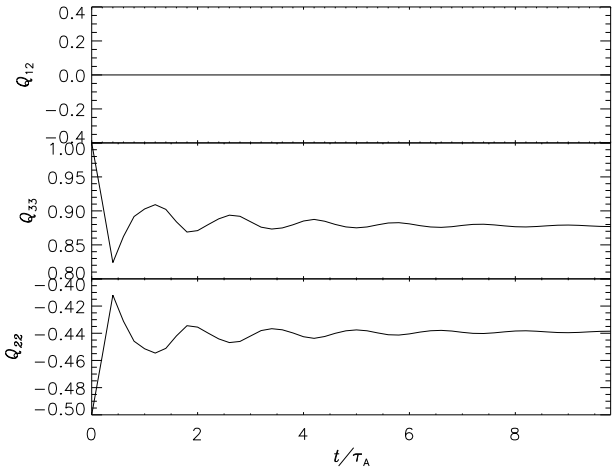}
  \caption{Mass quadrupole moments for model B, normalised to the maximum of $Q_{33}$
    ($1.3\times10^{25}$ g cm$^2$) as a function of time in units of the
    Alfv\'{e}n time. We find $Q_{22}=-Q_{33}/2$ and $Q_{12}=0$, as expected
    for an axisymmetric configuration. The mountain performs damped
    lateral oscillations with twice the frequency of model A.}
  \label{fig:stab:modb_quadru}
\end{figure}

\change{
\subsection{Uniform toroidal field}
There are strong theoretical indications that the magnetorotational
instability \citep{Balbus98} acts during core collapse
supernova explosions to generate a substantial toroidal field
component $B_\phi \sim B_p$ beneath the stellar surface
\citep{Cutler2002, Akiyama03}.
The hydromagnetic stability of equilibria with $B_\phi \ne 0$ will be
discussed thoroughly in a forthcoming paper. In this subsection, for
completeness, we present the results of a preliminary investigation.
}

\change{
Let us rerun model A with the same initial conditions while applying a
uniform $B_\phi = 10^{-7} B_0 = 0.35 B_p$ throughout the integration volume and at
$r=R_\ast$, where the poloidal field component is defined as
$B_p=(B_r^2+B_\theta^2)^{1/2}$, taken at the point $[\tilde{x}=(r-R_\ast)/h_0,
\theta, \phi]=(10^{-3}, 0.012, 0)$. $B_\phi$ is uniform only initially and is allowed to
evolve nonuniformly as \textsc{zeus-mp} proceeds. This procedure leads to a non-equilibrium
configuration, because we do not generalise and solve again the GS
equation (\ref{eq:mountains:gs_master}) to accomodate $B_\phi\ne
0$. Nevertheless, it provides us with some insight into the
stability of a field with nonzero pitch angle.
}

\begin{figure}
\includegraphics[width=84mm, keepaspectratio]{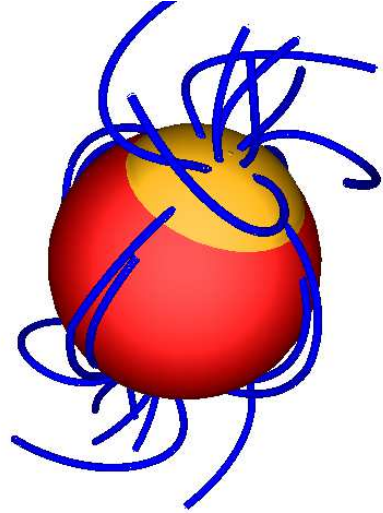}
\caption{\change{Model A repeated with the same parameters as in Table
    \ref{tab:models} including a
  uniform toroidal field $B_\phi = 10^{-7} B_0 = 0.35 B_p$. The
    mountain is  defined by the orange isosurface
    $\rho(r,\theta,\phi)=1.04\times10^{9}$ g cm$^3$, while red
      denotes the neutron star surface $r=R_\ast$. In order to
    improve visibility, all length scales of the mountain and the
    field lines (blue) are magnified five-fold. The field exhibits
  a helical topology, which is most distinct in the polar flux tubes where the
  poloidal contribution is weakest.}}
  \label{fig:stab:moda_constphi_3d}
\end{figure}

\begin{figure}
\includegraphics[width=84mm, keepaspectratio]{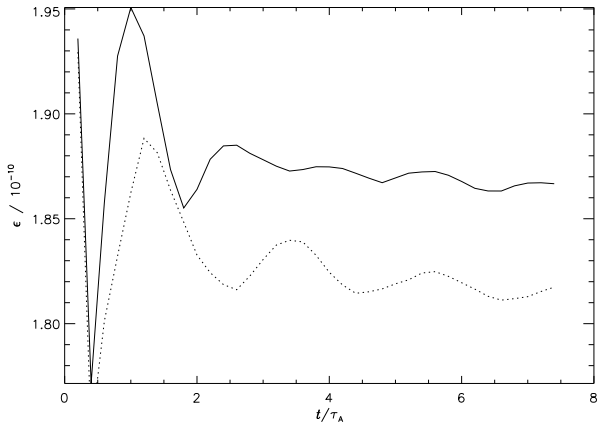}
\caption{\change{Mass ellipticity of model A with (solid) and without
    (dotted) a uniform $B_\phi$. The toroidal field component leads to
    a shorter oscillation period and to a higher saturation
    ellipticity.}}
    \label{fig:stab:moda_constphi_ellipticity}
\end{figure}

\change{
\fref{stab:moda_constphi_3d} displays the result of this numerical experiment after
$t=7.4 \tau_A$. The toroidal field component creates a helical field
topology in the polar region far from the surface, 
where the poloidal field is comparably weak. In the equatorial region,
however, the poloidal field is still dominant and the structure
remains unchanged from \fref{stab:moda2d}. Remarkably, the toroidal field does not alter the
stability of the system qualitatively, at least for the parameters of model A [We
expect a stronger effect in other parameter regimes; see
\citet{Lifschitz89,Goedbloed04}]. The ellipticity, displayed in
\fref{stab:moda_constphi_ellipticity} (solid curve), exhibits
characteristic oscillations with a period $\sim 30$ per cent smaller 
than that of the purely poloidal configuration (dotted curve), which
can be explained simply by the increase in the Alfven speed. In
addition, the saturation ellipticity is $\sim 3$ per
cent higher than in model A; the magnetic tension increases with $B$,
sustaining the mountain at a lower colatitude.
}
\section{Nonaxisymmetric stability}
\label{sec:stab3d}

We turn now to the three-dimensional evolution of a magnetised
mountain in \textsc{zeus-mp}. The chief finding, presented below, is
that the initial (axisymmetric) configuration becomes unstable to
toroidal perturbations, but that, after a brief transition phase, the
system settles into a new (nearly axisymmetric) state, which is stable
in the long term. Section \ref{sec:stab3d:reference} compares the
results to the axisymmetric reference model A. The magnetic and mass
multipole moments are computed in section \ref{sec:stab3d:moments},
the influence of the boundary conditions is considered in section
\ref{sec:stab3d:bc}, and the component-wise  evolution of the energy
is examined in section \ref{sec:stab3d:energy}. A scaling of the mass
quadrupole moment versus $M_a$ is derived empirically in section
\ref{sec:stab3d:massvar}. The curvature rescaling is verified
in section \ref{sec:stab3d:real}.

\subsection{General features}
\label{sec:stab3d:reference}

\begin{figure*}
  \includegraphics[width=168mm, keepaspectratio]{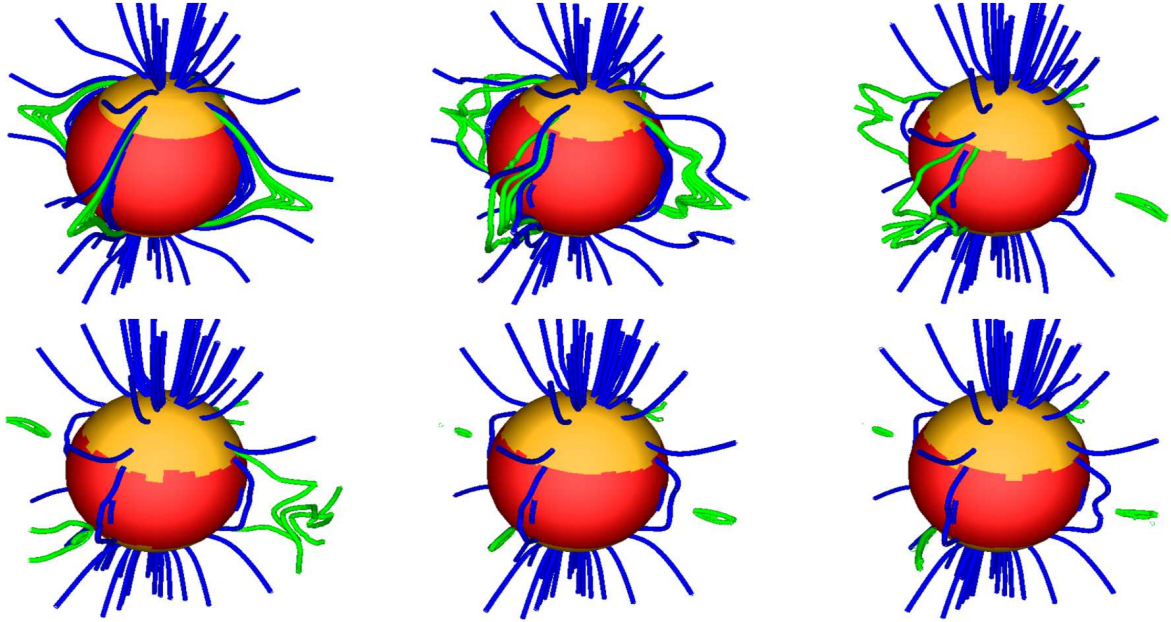}
  \caption{Density and magnetic field of model D at $\tau/\tau_\mathrm{A}=0, 1,
    2, 3, 4, 5$ (from top left to bottom right). The
    mountain is  defined by the orange isosurface
    $\rho(r,\theta,\phi)=1.04\times10^{9}$ g cm$^3$\change{, while red
      indicates the neutron star surface $r=R_\ast$}. In order to
    improve visibility, all length scales of the mountain and the
    field lines are magnified five-fold. The mountain becomes unstable to toroidal modes at $\tau \approx 0.8 \tau_\mathrm{A}$.
    It subsequently relaxes to a new nonaxisymmetric
    equilibrium. The footpoint of the blue fieldlines is at the 
    stellar surface while green fieldlines are traced starting from
    the equator. Green field lines eventually become topologically
    disconnected (see text).}  
  \label{fig:stab:modd3c}
\end{figure*}

\begin{figure}
  \includegraphics[width=84mm, keepaspectratio]{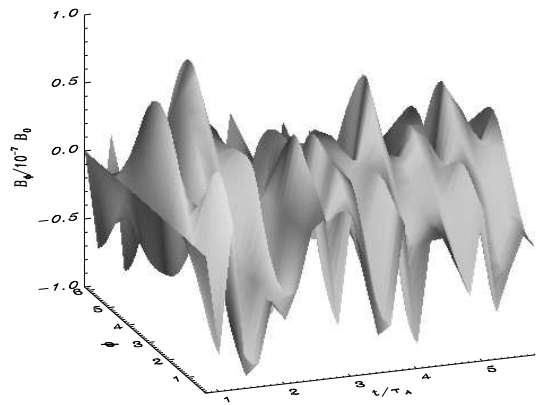}
  \caption{Evolution of the azimuthal magnetic field component $B_\phi$ at
    $(\tilde{x}, \theta)=(10^{-3}, 0.01)$ in model D as a function of longitude
    $\phi$ (in radians) and time $t$ (in units of the Alfv\'{e}n time.)}
  \label{fig:stab:modd_phiwave}
\end{figure}

Model D starts from the same configuration as model A ($M_a/M_c=1.0$,
\textrm{outflow} at $r=R_m$) but is evolved in three dimensions. Six snapshots of
a density isosurface (orange) and magnetic field lines (blue) are depicted in Fig.
\ref{fig:stab:modd3c}. At $\tau \approx 0.8 \tau_\mathrm{A} = 200 \tau_0$, the system 
undergoes a violent transition. The field lines bend in the $\phi$
direction, indicating that the initial axisymmetric configuration is
unstable to toroidal modes, a channel that is evidently not present in
axisymmetric simulations. This hypothesis is supported by
\fref{stab:modd_phiwave} which plots $B_\phi$ at $\tilde{x}=10^{-3}$ and
$\theta=0.01$ as a function of $\phi$ and $t$.  The magnetic
field takes the form of an azimuthal travelling wave $B_\phi \sim
\exp[\mathrm{i}(m \phi - \omega t)]$. From \fref{stab:modd_phiwave},
we measure the phase speed to be approximately $v_\mathrm{p}=\omega R_\ast/m=27
v_\mathrm{A}$, where the Alfv\'{e}n speed $v_\mathrm{A}$ is measured at $(\tilde{x},
\theta, \phi)=(10^{-3}, 0.01, 0.1)$ and we assume
$m=1$. An inhomogenous plasma generally supports mixed
magnetosonic/Alfv\'{e}n modes, so $v_\phi$ does not
necessarily equal $v_\mathrm{A}$ or the fast/slow magnetosonic
speed. The magnitude of $B_\phi$ is comparable to the magnitude of the
polar magnetic field $B_\mathrm{p}=2.9 \times 10^{-7} B_0$.

The deviations of the mountain isosurface, defined by $\rho(r, \theta,
\phi)=1.04 \times 10^9$ g cm$^3$, from axisymmetry are small ($\la 5$ per
cent laterally and $\la 0.06$ per cent radially during the transition
phase). The isosurface spreads outward by $\sim 32$ per cent relative
to ites initial position.

\begin{figure}
  \includegraphics[width=84mm, keepaspectratio]{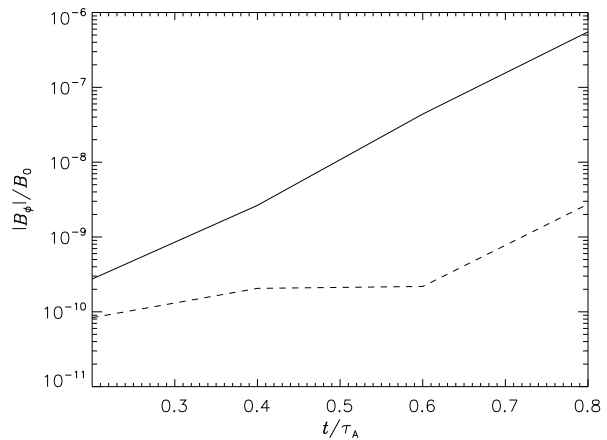}
  \caption{$|B_\phi|$ at $(\tilde{x},\theta, \phi)=(10^{-3}, 1.5, 1.3)$ for
    model D, simulated at higher resolution ($N_\phi=32$, solid) and lower
    resolution ($N_\phi=8$, dashed). Small wavelength perturbations
    grow faster. We track the absolute value of $|B_\phi|$
    in order to isolate better the dominant mode, as the system exists in a
    superposition of stable and unstable modes, and $B_\phi$ switches sign.}
  \label{fig:stab:growthrate}
\end{figure}

\fref{stab:growthrate} demonstrates how the instability grows. We plot
$|B_\phi|$ at the (arbitrary) position $(r,
\theta, \phi)=(R_\ast, 1.5, 1.3)$ versus time. The solid
curve corresponds to a higher
toroidal resolution ($N_\phi=32$ grid cells in $\phi$ direction) than
the dashed curve ($N_\phi=8$). We note first that $|B_\phi|$ grows
exponentially with time, as expected in the linear regime. The growth
rate is measured to be $\Gamma=\mathrm{Im}(\omega)=0.05
\tau_0^{-1}=12.5 \tau_\mathrm{A}^{-1}$. Second, the
instability is manifestly associated with toroidal
modes. The magnetic perturbation, $\delta
\mathbf{B}$, induced by a linear Lagrangian displacement $\bxi$ is
$\delta \mathbf{B} = \nabla \times (\bxi \times \mathbf{B})$.
By writing out the vector components, one sees that $\delta
B_\phi \ne 0$ implies $\xi_\phi \ne 0$, provided the unperturbed field
has the form $\mathbf{B}=B_r(r, \theta) \hat{\mathbf{e}}_r + B_\theta(r,\theta)
\hat{\mathbf{e}}_\theta$.

The dashed curve in \fref{stab:growthrate} tracks $|B_\phi|$ for a
simulation carried out at a lower resolution ($N_\phi=8$). The
instability grows significantly slower with
$\Gamma=\mathrm{Im}(\omega)=0.02 \tau_0^{-1}=5 \tau_\mathrm{A}^{-1}$. We
conclude that $\Gamma$ scales with the wavelength $\lambda$ of the
perturbation roughly as $\lambda^{-1/2}$. The piecewise-straight appearance of
the dashed curve in \fref{stab:growthrate} shows that the global
oscillations are governed by a superposition of unstable (growing) and
stable wave modes.

What type of instability is at work here? In order to answer that
question, we first write down the change in potential energy associated
with a Lagrangian displacement $\bxi$ in a form that reveals the physical
meaning of the different contributions \citep{Biskamp93,Lifschitz89,Greene68}:
\begin{eqnarray}
  \delta W_p & = & \frac{1}{2} \int \mathrm{d}V
  \left[ |\mathbf{Q}_\perp|^2+ |\nabla \bcdot \bxi_\perp + 2 \bkappa
    \bcdot \bxi_\perp|^2 B_0^2 \right. \nonumber \\
  & & +c_s^2 \rho |\nabla \bcdot \bxi|^2 - j_\parallel
  (\bxi_\perp^\ast \times \mathbf{b})\bcdot \mathbf{Q} \nonumber \\
  & & \left.  - 2(\bxi_\perp \bcdot \nabla p)(\bkappa \bcdot \bxi_\perp^\ast)
  - (\bxi^\ast \bcdot \nabla \varphi) \nabla \bcdot (\rho \bxi)\right],
\end{eqnarray}
We include the term due to gravity \citep{Goedbloed04} and define
$\mathbf{j}=\mu_0^{-1} \nabla \times \mathbf{B}$ (current density),
$\mathbf{Q}=\nabla \times (\bxi \times \mathbf{B})$ (change in $\mathbf{B}$ as a response to
$\bxi$), $\bkappa=(\mathbf{b}\bcdot \nabla)\mathbf{b}$ (field line
curvature), and $\mathbf{b}=\mathbf{B}/B$. The
subscripts $\perp$, $\parallel$ refer to the magnetic field, such that
$\balpha_\perp=\balpha-(\balpha \bcdot \mathbf{b})\mathbf{b}$ and
$\alpha_\parallel=\balpha \bcdot \mathbf{b}$. The first three
(stabilising) terms are the potential energy of the shear Alfv\'{e}n
mode, the fast magnetosonic mode, and the (unmagnetized) sound
mode. They are all positive definite. The last three terms may have either sign. The term
proportional to $j_\parallel$ causes the current-driven instabilities,
the curvature term causes pressure-driven instabilities (when $\bkappa \bcdot
\nabla \rho >0$), and the final term causes gravitational instabilities.

We first note that $j_\parallel=0$, ruling out current-driven
instabilities. Furthermore, we have $\bkappa=\kappa_r(r,\theta) \hat{\mathbf{e}}_r +
\kappa_\theta(r, \theta) \hat{\mathbf{e}}_\theta$ in our particular
field  geometry. In principle, this term admits pressure-driven
instabilities. However, we would expect such instabilities, if they
exist, to also grow in an axisymmetric system, yet they do not. This
suggests that the instability we see in
Figs. \ref{fig:stab:modd3c}--\ref{fig:stab:growthrate} is associated with a toroidal
dependence in $\bxi$, leaving the gravitational term, which indeed
contains $\partial_\phi \bxi$ contributions.

One prominent gravitational mode is the Parker
or magnetic buoyancy instability \citep[PM04;][]{Mouschovias74}. Its physics was elucidated by \citet{Hughes87} for
 a plane-parallel, stratified atmosphere with a horizontal field
increasing with depth $z$. The instability involves an interchange
sub-mode and an undular sub-mode. The interchange sub-mode satisfies
$k_y=\xi_y=0$. We do not observe this mode in our system because (i)
it should also be present in two dimensions, as it does not rely on a
toroidal dependence, yet it is absent; and (ii) it is
inconsistent with the line-tying boundary condition at $r=R_\ast$.
On the other hand, undular modes compress the plasma along
field lines, even in systems which are
interchange stable. In two dimensions, they are
restricted to $k_x=Q_x=0$, whereas a non-vanishing $Q_x$ is allowed in
three dimensions. \citet{Hughes87} showed that
$\delta W_p$ is minimized for $k_x \rightarrow \infty$,
consistent with the results in \fref{stab:growthrate};
the instability grows faster, if we
allow smaller wavelength perturbations by increasing $N_\phi$.

When $v_A$ is uniform the growth rate of the Parker instability
reaches an asymptotic maximum $\Gamma_\mathrm{P} \simeq (g/\Lambda)^{1/2}$ for $k_x
\Lambda, k_y \Lambda \gg 1$. Here, $\Lambda=v_\mathrm{A}^2/g$ is the scale height for a
stratified atmosphere with uniform gravitational acceleration
$g$. We recognize $(\Lambda/g)^{1/2}$ as the characteristic free fall
time over one scale height. In the units specified in
\sref{model:units}, we find $\Gamma_\mathrm{P} \tau_0 \simeq \Lambda \simeq g
\simeq 1$, two orders of magnitude higher than the observed
growth rate $\Gamma \simeq 10^{-2} \tau_0^{-1}$. The discrepancy
arises because the Parker instability cannot grow freely in the belt
region, since the adjacent plasma at higher latitudes effectively acts
as a line-tying boundary for the magnetic field.

The snapshot at $t = \tau_A$
(top-middle panel in \fref{stab:modd3c}) demonstrates how the
instability starts in the equatorial region, whose magnetic belt
represents the endpoint of the two-dimensional Parker
instability. The undular Parker sub-mode  releases
gravitational energy by radial plasma flow towards the neutron star's
surface. At the same time, the magnetic field is rearranged such as to minimize the
(radial) gradient in $\mathbf{B}$. Importantly, the undular mode is
not available in the axisymmetric case. The extra degree of freedom in
the $\phi$ direction allows perturbations to develop which do no work
against the magnetic pressure, destabilising the belt region.

Particularly interesting here is the formation of topologically
disconnected field lines (green curves in \fref{stab:modd3c}). These occur
when field lines are pushed out of the radial boundary surface. They
are then disrupted and can subsequently reconnect at the equatorial
boundary, forming O-type neutral points (``bubbles'') and associated
Y-type points. It is important to note, however, that the formation of
these bubbles is not an unphysical boundary effect. Instead, in a
realistic setting, even a small resistivity leads to reconnection
and thus to a topological rearrangement of the field. The
effect is similar to the formation of plasmoids
\citep{Schindler88}. $B_r$ switches sign at the magnetic equator,
implying the existence of a current sheet. Reconnection then leads to the creation
of magnetic X-type neutral points and the associated bubbles. We
discuss resistive effects and the importance of these bubbles to
resistive instabilities in an accompanying paper.

\begin{figure*}
  \includegraphics[width=80mm, keepaspectratio]{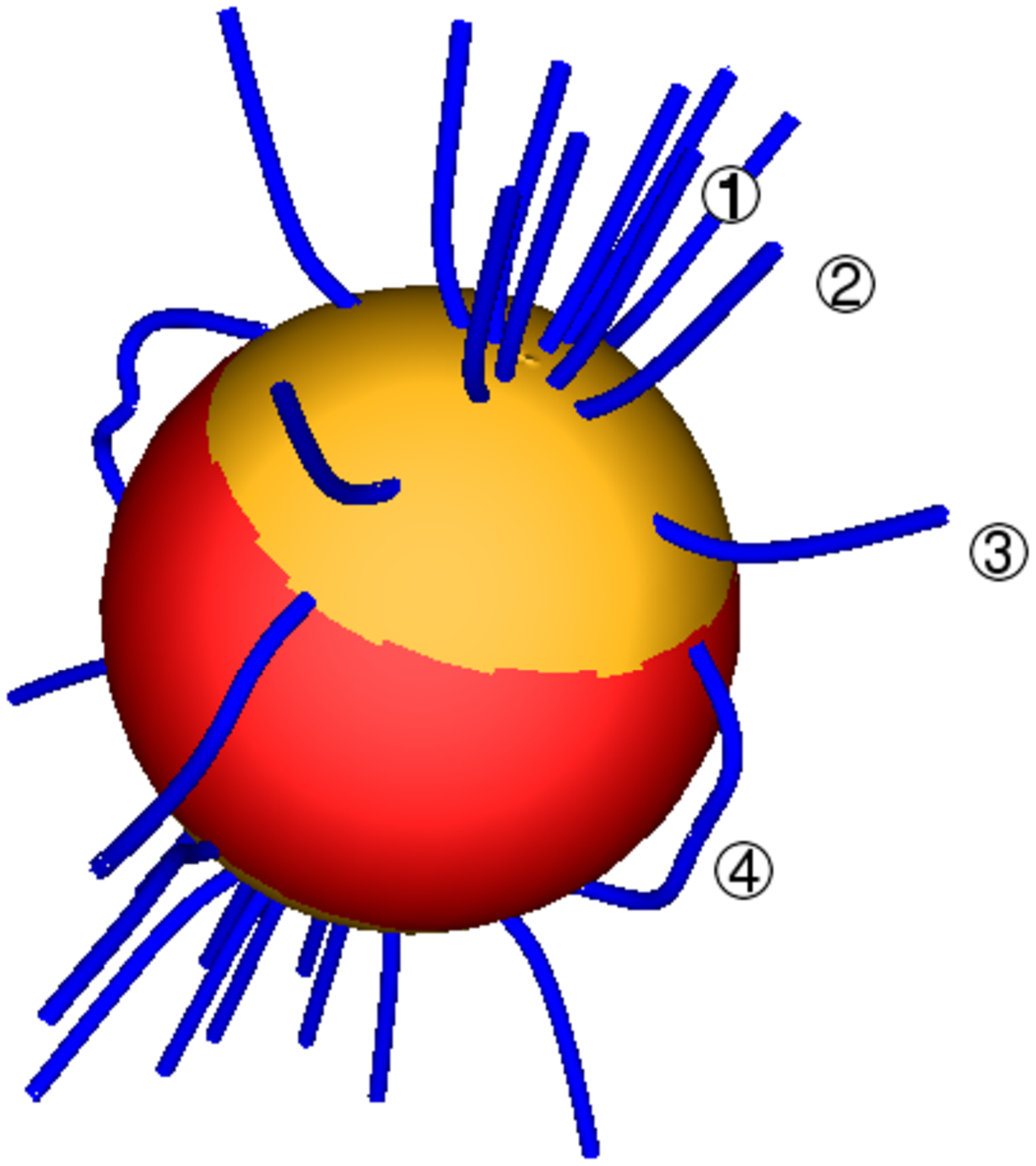}
  \includegraphics[width=80mm, keepaspectratio]{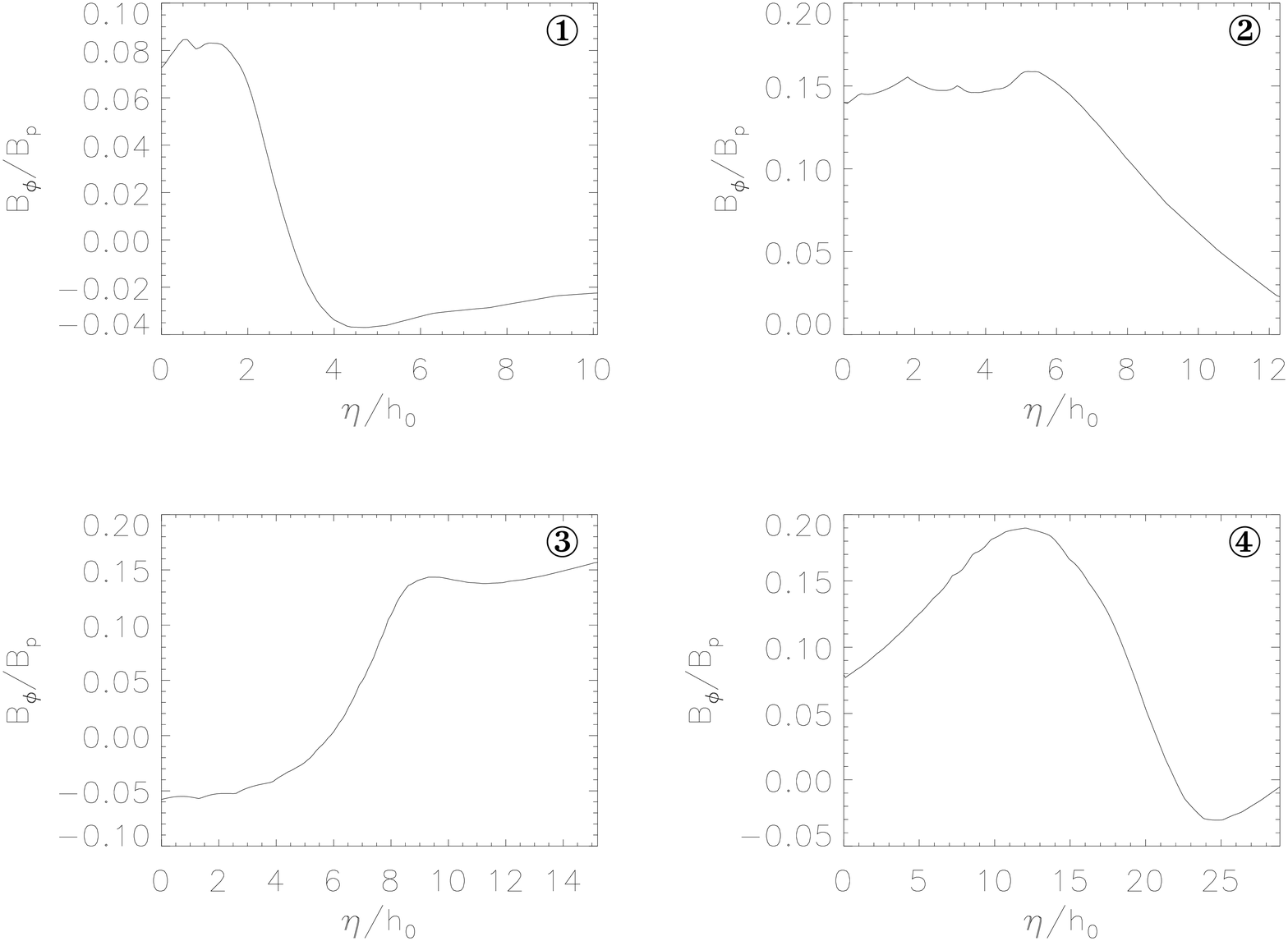}
  \caption{Magnetic pitch angle $B_\phi/B_p$ (right panel) as a function of the
    coordinate \change{$\eta$} along four magnetic field lines 1--4
    for model D, for a snapshot taken at $t=5 \tau_\mathrm{A}$. The
    positions of the field lines are depicted in the left panel. The
    mountain is  defined by the orange isosurface
    $\rho(r,\theta,\phi)=1.04\times10^{9}$ g cm$^3$, while red
      denotes the neutron star surface $r=R_\ast$. In order to
    improve visibility, all length scales of the mountain and the
    field lines (blue) are magnified five-fold.}
  \label{fig:stab:pitch_angle}
\end{figure*}

The concept of rational magnetic surfaces,
where the field lines close upon themselves, plays an important role
in a local plasma stability analysis. The bending of field lines as a result
of a Lagrangian displacement is associated with an increase in
potential energy. Hence, for almost all instabilities to occur, this
contribution, which can be expressed as $\mathbf{B} \bcdot \nabla
\bxi$, needs to be small. In a tokamak geometry, it can be shown that
this term vanishes on a rational surface. The spatial location of
rational surfaces is directly related to the 
pitch angle $B_\phi/B_p$. We defer a detailed
analysis of the rational magnetic surfaces in our problem to a
forthcoming paper and restrict ourselves to a brief discussion in the
following paragraph.

In \fref{stab:pitch_angle}, we plot the pitch angle as a function of
the coordinate $\eta$ along the field line (right panel) for four
different field lines (left panel) in model D. We first note that the
toroidal component stays below 20 per cent of the poloidal component
along all four field lines. Furthermore, the absolute magnitude of the pitch angle
tends to increase with colatitude. This is consistent with the
previous discussion. The Parker instability (and hence
$B_\phi$) dominates close to the magnetic
equator. The wave-like character of the instability is vividly
demonstrated by the zero crossings of the pitch angle.

\subsection{Mass and magnetic quadrupole moments}
\label{sec:stab3d:moments}
\begin{figure}
  \includegraphics[width=84mm, keepaspectratio]{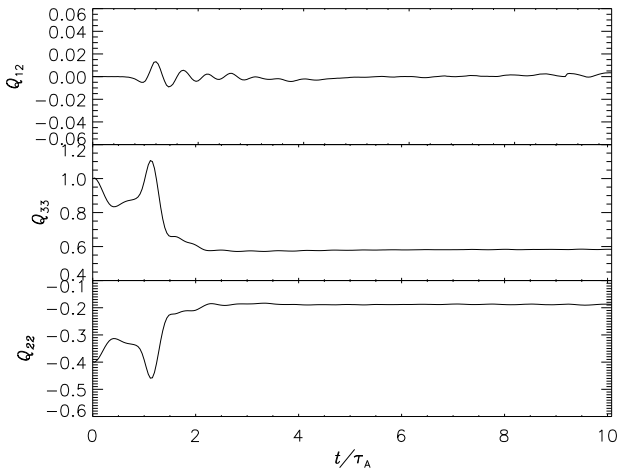}
  \caption{Mass quadrupole moments for model D, normalised to the maximum of
    $Q_{33}$ ($1.30\times10^{25}$ g cm$^2$) as a function of time in
    units of the Alfv\'{e}n time. The sytem develops a
    substantial asymmetry, characterised by the off-diagonal element
    $Q_{12}$, during the relaxation phase, before settling down to a
    nearly axisymmetric state.}
  \label{fig:stab:modd_quadru}
\end{figure}

At $t \approx 2 \tau_\mathrm{A}$, the system in \fref{stab:modd3c} settles down to a stable
state which differs from the initial configuration, primarily by being
nonaxisymmetric with respect to the pre-accretion magnetic axis. The
field lines whose footpoints are at a low colatitude move
towards the magnetic poles. This behaviour is reflected in the mass quadrupole moments,
plotted against time in \fref{stab:modd_quadru}. The transition to a
nonaxisymmetric magnetic field configuration at $t \approx  \tau_A$
is accompanied by a sudden rise in the off-diagonal moment
$Q_{12}$. However, by the time the mountain settles down at $t \approx
2 \tau_A$, axisymmetry is largely restored and $Q_{12}$ decreases.
\fref{stab:modd_quadru} shows that $Q_{12}$ 
oscillates before damping down,
with a remarkably low deviation from axisymmetry of $Q_{12}/Q_{33} <
0.1$ per cent in the final state.

We reiterate that the final state is not the same as the initial
state, even though it is nearly axisymmetric. Furthermore, the final
state is stable. This is the main result of the paper, as far as
astrophysical applications are concerned.

\begin{figure*}
  \includegraphics[width=140mm, keepaspectratio]{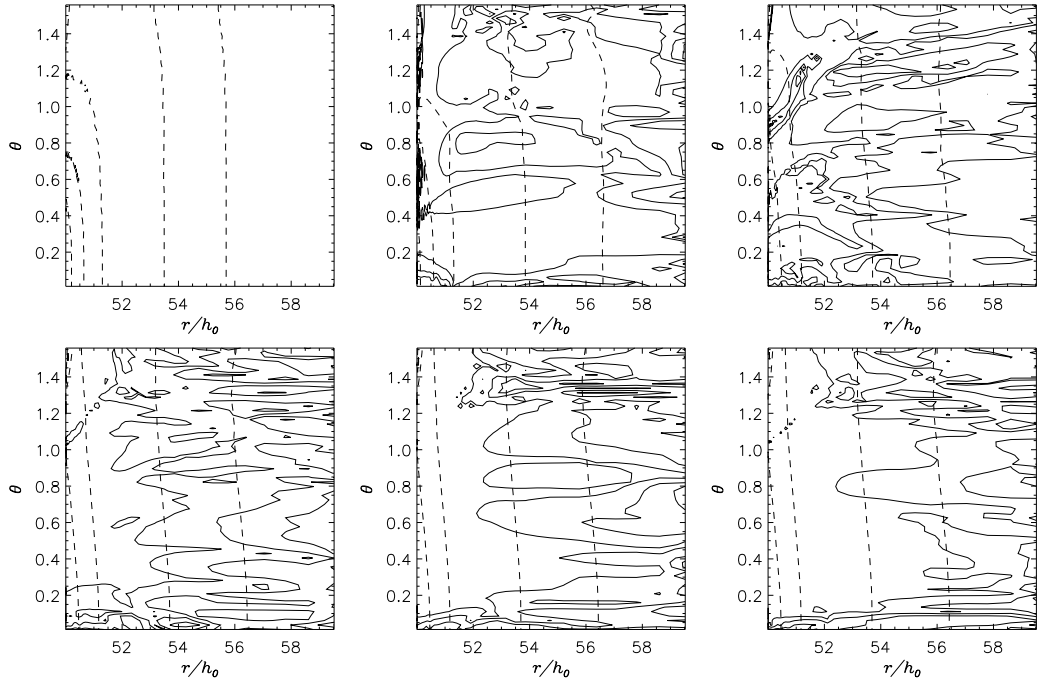}
  \caption{Meridional section for model D at $t/\tau_A=0, 1, 2, 3, 4,
    5$  (top left to bottom right). Shown are the density contours
    (dashed curves) for $\log_{10}(\rho/\rho_0')=-13, -12, -11, -10.7,
    -10.5, -10.3$, and the normalised Lorentz force per unit volume
    $(\mathbf{J} \times \mathbf{B})_\phi /|\mathbf{J} \times
    \mathbf{B}|$ (solid curves) for the values 0.1, 0.5, 
      0.9. The Lorentz force develops a toroidal component as $B_\phi$
      increases, but its poloidal component diminishes, allowing the
      poloidal pressure gradient to push the mountain equatorwards.}
  \label{fig:stab:modd2d}
\end{figure*}

Why does $Q_{ij}$ decrease? Naively, one
would not expect a significant change, given that the Parker
instability predominantly acts in the equatorial belt region, while most of
the plasma is located at the magnetic pole. The answer can be found in
the Lorentz force, which balances the lateral pressure
gradient. \fref{stab:modd2d} shows how the 
relative strength of the toroidal component of the Lorentz force,
$(\mathbf{J} \times \mathbf{B})_\phi /|\mathbf{J} \times 
\mathbf{B}|$ (dashed curve), grows as a function of time in
model D. As $B_\phi$ grows, following the onset
of the instability, the force per unit volume develops a toroidal component
while its lateral component decreases. Hence the (approximately
unchanged) lateral hydrostatic pressure gradient
forces the mountain to slip towards the equator. After the
system settles down, $B_\phi$ decreases and the lateral components of $\mathbf{J} \times
\mathbf{B}$ and $\nabla p$ readjust to balance each other, leading to
the stable equilibrium state.

The ellipticity $\epsilon \propto Q_{22}$ reaches a local maximum during the
transition phase at $t \approx \tau_A$ and subsequently drops. The
mass quadrupole moment of the final configuration is $\approx 33$ per cent
lower than in model A. The asymptotic values of $Q_{ij}$ for the eight
models in Table \ref{tab:models} are tabulated in Table
\ref{tab:quad}, normalized to $Q_{33}(t=8 \tau_\mathrm{A})$ for model A.

\begin{table}
  \centering
  \caption{Asymptotic values of $Q_{ij}$ for the eight models in Table
    \ref{tab:models}, normalized to $\hat{Q}_{33}=Q_{33}(t=8 \tau_\mathrm{A})$ for
    model A. We select $t=8 \tau_\mathrm{A}$ (models A--E), $t=5
    \tau_\mathrm{A}$ (models F \& G), and $t=4 \tau_\mathrm{A}$ (models
  J \& K) to compute the asymptotic value.}
  \begin{tabular}{@{}cccc}
    \hline
    Model & $Q_{12}/\hat{Q}_{33}$ & $Q_{22}/\hat{Q}_{33}$ & $Q_{33}/\hat{Q}_{33}$ \\ \hline
    A & $-3.1 \times 10^{-8}$ & $-0.50$ & $1.00$ \\
    B & $-3.1 \times 10^{-8}$ & $-0.51$ & $1.02$ \\ \hline
    D & $1.6 \times 10^{-3} $ & $-0.21$ & $0.64$ \\
    E & $-3.1 \times 10^{-8}$ & $-0.51$ & $1.02$ \\ \hline
    F & $-1.9 \times 10^{-3}$ & $-0.23$ & $0.60$ \\
    G & $8.7 \times 10^{-4}$ & $-0.17$ & $0.68$ \\ \hline
    J & $5.7 \times 10^{-2}$ & $-1.27$ & $3.9$ \\
    K & $2.5 \times 10^{-4}$ & $-4.72$ & $12$ \\ \hline
  \end{tabular}
  \label{tab:quad}
\end{table}

\begin{figure}
  \includegraphics[width=84mm, keepaspectratio]{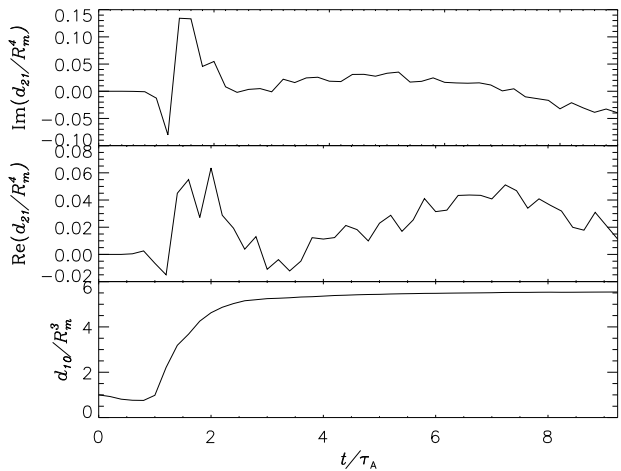}
  \caption{Magnetic dipole moment $d_{10}/R_m^3$ (bottom) and magnetic quadrupole
    moment $d_{21}/R_m^4$ (top, middle) for model D, normalised to the initial value of
    \change{$d_{10}/R_m^3=5.29\times10^{-7} B_0$}, as a function of time (in units
    of the Alfv\'{e}n time scale). All other components vanish due to symmetry.}
  \label{fig:stab:modd_magmoments}
\end{figure}

The nonvanishing components of the magnetic dipole and quadrupole moments $d_{lm}(r=R_m)$, defined as
\begin{equation}
    d_{lm}(r=R_m) = R_m^{l+2} \int \mathrm{d}\Omega\; Y_{lm}^\ast B_r
\end{equation}
(see appendix \ref{sec:app:magmulti}), are displayed as functions of
time in \fref{stab:modd_quadru}. The
dipole moment $d_{10} =4 (\pi/3)^{1/2} \mu$ increases rapidly during the transition phase,
reaching an asymptotic maximum of 5.5 times the initial
value. Likewise, the quadrupole \change{$d_{21}$} peaks during the
transition phase before settling down to a constant value. The final
field is highly axisymmetric, deviating from perfect symmetry by $|d_{21}|
R_m/d_{10}=0.8$ per cent. The asymptotic values of $d_{lm}$ for models
A--K are listed in Table \ref{tab:magnetic}.

\begin{table}
  \centering
  \caption{Non-vanishing components of the asymptotic magnetic dipole
    moments $d_{10}/R_m^3$ and magnetic quadrupole moments
    $q_{10}/R_m^4$, both normalised to $\hat{d}_{10}/R_m^3=d_{10}(t=8
    \tau_\mathrm{A})/R_m^3$. We select $t=8 \tau_\mathrm{A}$ (models A--E), $t=5
    \tau_\mathrm{A}$ (models F \& G), and $t=4 \tau_\mathrm{A}$ (models
  J \& K) to compute the asymptotic value.}
  \begin{tabular}{@{}cccc}
    \hline
    Model & $d_{10}/\hat{d}_{10}$ & $\mathrm{Re}(d_{21})/(\hat{d}_{10}
    R_m) $ & $\mathrm{Im}(d_{21})/(\hat{d}_{10} R_m)$ \\ \hline
    A & $1.0$ & $-4.6 \times 10^{-8}$ & $-4.8 \times 10^{-9}$ \\
    B & $1.3$ & $-5.9 \times 10^{-8}$ & $-6.1 \times 10^{-9}$ \\ \hline
    D & $7.4$ & $4.8 \times 10^{-2}$ & $-4.3 \times 10^{-2}$ \\
    E & $1.3$ & $-5.9 \times 10^{-8}$ & $-6.1 \times 10^{-9}$ \\ \hline
    F & $7.3$ & $1.1 \times 10^{-1}$ & $8.2 \times 10^{-3}$ \\
    G & $7.1$ & $2.1 \times 10^{-2}$ & $-5.5 \times 10^{-2}$ \\ \hline
    J & $21$ & $-1.2$ & $-0.12$ \\
    K & $57$ & $-6.3 \times 10^{-4}$ & $9.1 \times 10^{-4}$ \\ \hline
  \end{tabular}
  \label{tab:magnetic}
\end{table}

\subsection{Boundary conditions}
\label{sec:stab3d:bc}
\begin{figure}
  \includegraphics[width=84mm, keepaspectratio]{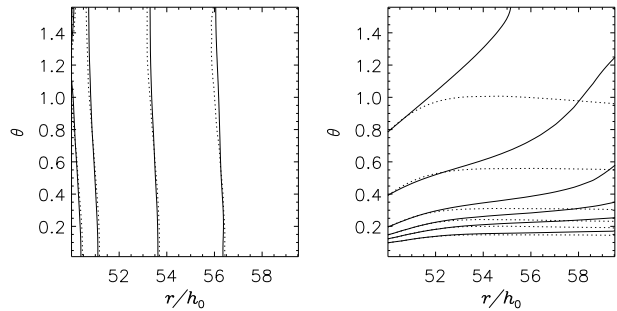}
  \caption{Meridional section of density contours (left) and magnetic
    field lines (right) for model D (solid curve) and model E (dotted
    curve). The \texttt{inflow} boundary condition corresponds to line-tying at the outer
    boundary. Deviations between the two models occur in the
    outermost, low density regions.}
  \label{fig:stab:mode3d}
\end{figure}

We perform a simulation (model E) with the same initial configuration as model D
($M_a/M_c=1.0$ and $b=3$) but with \texttt{inflow} boundary conditions
at $R=R_m$. \fref{stab:mode3d} compares the density (left panel) and magnetic
field (right panel) of models D and E. Again, 
\texttt{inflow} pins the magnetic field at the outer
boundary, as opposed to \texttt{outflow}, which leaves the field
free. The density distribution is almost unaffected.
The magnetic field is mainly affected in the outermost region, where the
plasma density is low. The overall time evolution (a nonaxisymmetric
transition phase which leads to a nearly axisymmetric equilibrium)
remains as before, too. We therefore conclude that the outer boundary
condition can be chosen opportunistically.

By contrast, the inner boundary condition contributes fundamentally to
stability. The tension of the magnetic field, which is tied to
the stellar surface, suppresses those modes which are driven by a
pressure gradient perpendicular to the magnetic flux surfaces, such as
the interchange and ballooning mode. If line-tying is taken away, the
latter modes disrupt the mountain in short order. If we rerun model A
(for example) by applying a \texttt{reflecting} boundary condition at
$r=R_\ast$, the mountain rapidly dissolves on a timescale
$\sim\tau_0$. The same experiment for model D results in high
velocities and steep field gradients, causing the numerical algorithm
of \textsc{zeus-mp} to break down.

\subsection{Energetics}
\label{sec:stab3d:energy}

\citet{Mouschovias74} showed that an isothermal, gravitating, MHD
system possesses a total energy $W$, which can be written as the sum of
gravitational, kinetic, magnetic, and acoustic contributions, defined
by the following volume integrals\change{, evaluated over the
  simulation volume}:
\begin{equation}
  \label{eq:stab:energy1}
  W_\mathrm{g}=\int \mathrm{d}V \rho \varphi,
\end{equation}
\begin{equation}
  W_\mathrm{k}=\frac{1}{2} \int \mathrm{d}V \rho v^2,
\end{equation}
\begin{equation}
  W_\mathrm{m}=\frac{1}{2 \mu_0}\int \mathrm{d}V B^2,
\end{equation}
\begin{equation}
  \label{eq:stab:energy4}
  W_\mathrm{a}=\int \mathrm{d}V p \log p.
\end{equation}
Here, $\mathbf{v}$ is the plasma velocity and $p=c_s^2 \rho$ is the
pressure.

\begin{figure}
  \includegraphics[width=84mm, keepaspectratio]{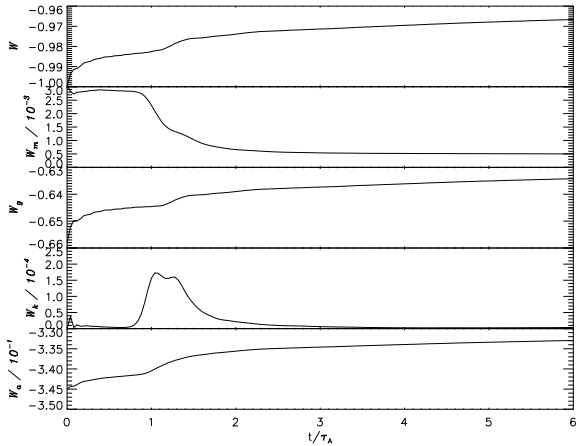}
  \caption{The evolution of the total energy $W$ and its components
    $W_\mathrm{m}$, $W_\mathrm{g}$,  $W_\mathrm{k}$, 
    and $W_\mathrm{a}$ (top to bottom) for model D, all
    normalised to $W_0=2.2\times10^{36}$ erg, as a function of time
    (in units of the Alfv\'{e}n time scale). The total energy
    increases artifically, due to mass loss through the outer
    boundary (see text).}
  \label{fig:stab:modd_energy}
\end{figure}

The evolution of \eeref{stab:energy1}--\eeref{stab:energy4} for model D is shown in
\fref{stab:modd_energy}. The magnetic energy (second panel from top)
steadily decreases to $20$ per cent of its
original value, as the axisymmetric equilibrium evolves to a lower
energy, nonaxisymmetric state. The kinetic energy peaks at $t=1.2
\tau_A$, during the transition phase when the magnetic reconfiguration
occurs. However, the gravitational and acoustic contributions, which
dominate $W$, increase with time. The reason for this
becomes apparent if we track the total mass in the simulation
volume. Approximately 3.7 per cent of the mass is lost through the
outflow boundary at $r=R_m$ by $t=6 \tau_\mathrm{A}$. The mass loss
is responsible for the increase of $W_\mathrm{a} \propto \rho^2$ and
$W_\mathrm{g} \propto \rho$, both of which are negative ($W_\mathrm{g}$ because
the plasma is gravitationally bound and $W_\mathrm{a}$ since $\rho <
1$ in our units).

\begin{figure}
  \includegraphics[width=84mm, keepaspectratio]{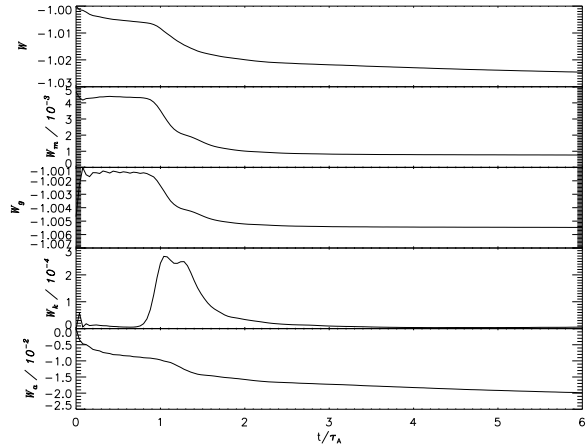}
  \caption{The evolution of the total energy $W$ and its components
    $W_\mathrm{m}$, $W_\mathrm{g}$,  $W_\mathrm{k}$, 
    and $W_\mathrm{a}$ (top to bottom) for model D, all
    normalised to $W_0=2.2\times10^{36}$ erg, as a function of time
    (in units of the Alfv\'{e}n time scale). $W$, $W_\mathrm{g}$, and
    $W_\mathrm{a}$ are now corrected for the mass loss through the
    outer boundary (cf. \fref{stab:modd_energy}).}
  \label{fig:stab:modd_energy_corrected}
\end{figure}

Let us try to correct for the mass loss by
multiplying $W_g$, $W_k$, and $W_a$ by $M(t=0)/M(t)$, where $M(t)$ is the mass
in the simulation volume at time $t$. The result is presented in \fref{stab:modd_energy_corrected}. $W_g$ and
$W_a$ now decrease, and the total energy,
$W=W_\mathrm{g}+W_\mathrm{k}+W_\mathrm{m}+W_\mathrm{a}$, decreases by just 2.5 per cent.

\begin{figure}
  \includegraphics[width=84mm, keepaspectratio]{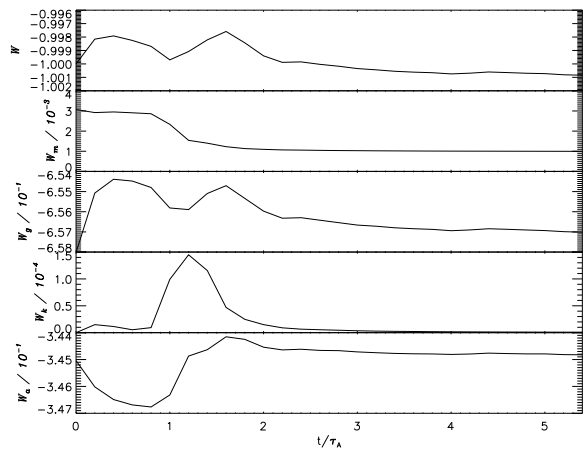}
  \caption{The evolution of the total energy $W$ and its components
    $W_\mathrm{m}$, $W_\mathrm{g}$,  $W_\mathrm{k}$, 
    and $W_\mathrm{a}$ (top to bottom) for model E, all
    normalised to $W_0=2.2\times10^{36}$ erg, as a function of time
    (in units of the Alfv\'{e}n time scale).}
  \label{fig:stab:mode_energy}
\end{figure}

The approximate correction above assumes $\rho$ decreases uniformly,
which is not strictly true. We therefore check
our claim that mass loss is responsible by tracking the energy evolution of model E, which has
the same initial configuration as model D, but an \texttt{inflow}
outer boundary which blocks mass loss. From \fref{stab:mode_energy},
it is clear that the total energy rises then falls,
consistent with the observed dynamical evolution. The mountain
oscillates until toroidal modes grow sufficiently to disrupt the
initial configuration and force it into a nonaxisymmetric state. There
is no spurious increase in $W$. We conclude that mass loss through the
outer boundary is indeed responsible for the observed behaviour of $W$
in model D in \fref{stab:modd_energy}.

\subsection{Dependence on  $M_a$}
\label{sec:stab3d:massvar}

\begin{figure}
  \includegraphics[width=84mm, keepaspectratio]{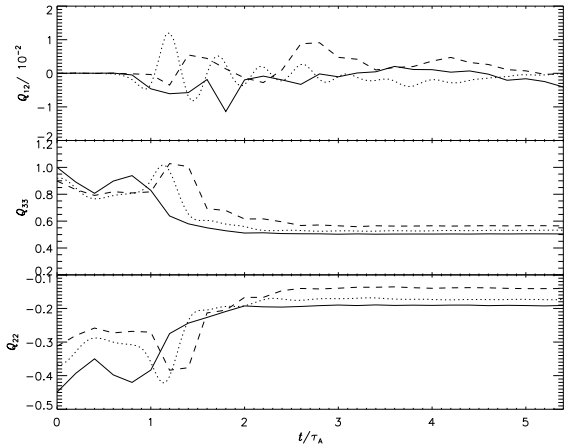}
  \caption{The mass quadrupole moments for models F (solid), D
    (dotted), and G (dashed), normalised to
    $1.33\times10^{25}$ g cm$^2$, as a function of time. While all models
    show  similar dynamical behaviour, the quadrupole moment of the
    final state increases with $M_a$.}
  \label{fig:stab:modmassvar_quadru}
\end{figure}
Does the final, nonaxisymmetric configuration of the mountain become
unstable once the accreted mass exceeds a critical threshold? There
are two ways that this can happen. First, the sequence of
nonaxisymmetric GS equilibria passed through as $M_a$ increases
can terminate above a critical value of $M_a$; i.e. there is a loss of
equilibrium. PM07 observed this phenomenon in axisymmetric magnetic
mountains with $M_a \ga 10^{-4} M_\odot$, when the source term in the
GS equation forces the flux function outside the range $0 \le \psi
\le \psi_\ast$ permitted by the boundary condition at
$r=R_\ast$. Second, the nonaxisymmetric state reached in
\fref{stab:modd3c} (for example) may be metastable. That is, it may be
a local energy minimum which can be reached from an axisymmetric
starting point via the Parker instability but which the system can
exit (in favor of some other, global energy minimum) if the system is
kicked hard enough. One way to kick the system hard is to increase
$M_a$ substantially.

We are not really in a position to answer this question definitively,
because the GS fails to converge to valid equilibria for $M_a \gg
10^{-4} M_\odot$, due to numerical difficulties (steep gradients,
which would be smoothed in a more realistic, non-ideal-MHD
simulation). Nevertheless, we begin to address the issue by performing
two simulations, models F and G, with the same parameters as model D
but with lower and higher masses viz. $M_a/M_c=0.6$ and $M_a/M_c=1.4$ respectively.
The mass quadrupole moments are plotted versus time in \fref{stab:modmassvar_quadru}.
The solid and dashed curves are for models F and G respectively, with
model D (dotted curve) overplotted for comparison.

The dynamical behaviour of all three models is similar: a violent
transition phase which settles down to a
nonaxisymmetric state. However, the start of the transition phase,
defined as the instant where $Q_{33}$ is maximized, scales
roughly $\propto 0.5 M_a/M_c$ in units of $\tau_A$. Physically, this means that the onset
of the toroidal instability depends on $M_a$. We can understand the trend
in terms of the Parker instability (\sref{stab3d:reference}), whose
growth rate scales as $\Gamma_\mathrm{P} \propto
v_\mathrm{A}^{-1}$. By measuring $v_A$ at $\theta=\pi/2$ in
models D,F, and G, we find $v_A \propto M_a$ empirically and
$\Gamma_\mathrm{P}\propto M_a^{-1}$, consistent with the Parker scalings.

\begin{figure}
  \includegraphics[width=84mm, keepaspectratio]{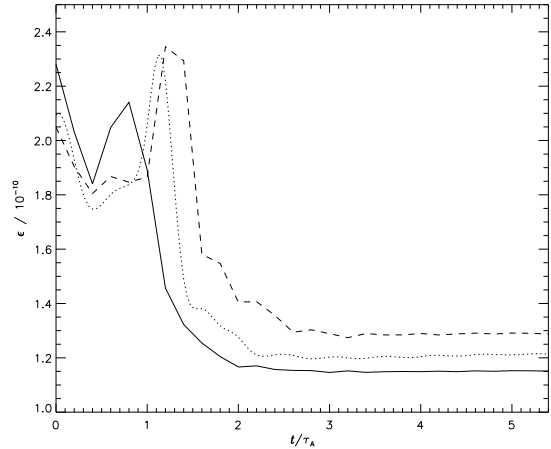}
  \caption{Ellipticity $\epsilon$ for models F (solid), D
    (dotted), and G (dashed) as a function of time. As expected,
    $\epsilon$ increases with $M_a$.}
  \label{fig:stab:modmassvar_ellipticity}
\end{figure}

The evolution of the ellipticity $\epsilon \propto Q_{33}$ for models
D, F, and G is displayed in \fref{stab:modmassvar_ellipticity}. Of
chief interest here is the ellipticity of the final state. It
increases along with $M_a$, consistent with \citet{Melatos05}. A
linear fit yields the following rule of thumb for our
\emph{downscaled} star (\sref{model:units}):

\begin{equation}
  \label{eq:stab:massscaling}
  \frac{\epsilon}{10^{-10}} = 1.12 \frac{M_a}{M_c}.
\end{equation}

Note, however, that the fit is valid in the range $0.6 \le M_a/M_c \le
1.4$. Numerical difficulties prevent us from extending it to larger
values of $M_a$. \citet{Payne06a} found $\epsilon/10^{-10} = 7.8
M_a/M_c (1+1.1M_a/M_c)^{-1}$  in the $M_a \sim M_c$ regime for the
axisymmetric equilibrium. Equation \eeref{stab:massscaling} yields
values roughly 70 per cent lower than the latter formula.

\begin{figure}
  \includegraphics[width=84mm, keepaspectratio]{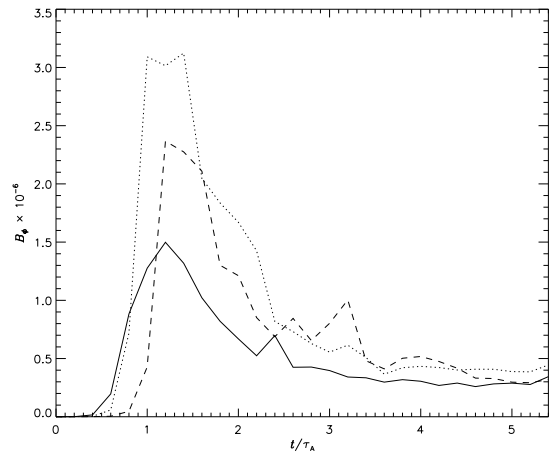}
  \caption{Azimuthal magnetic field
    component $|B_\phi|$ for models F (solid), D
    (dotted), and G (dashed), in units of $B_0$, plotted as a function
    of time, in units of the Alfv\'{e}n time.}
  \label{fig:stab:bphi}
\end{figure}

For completeness, we plot the magnitude of the toroidal field
component $|B_\phi|$ versus time in \fref{stab:bphi}. Interestingly,
the peak value is achieved for the intermediate mass model, D, not for
model G. However, $B_\phi$ in the final state depends weakly on $M_a$. We find
$B_{\phi,\mathrm{F}}=3.4 \times 10^{-7} B_0$, $B_{\phi,\mathrm{D}}=4.5
\times 10^{-7} B_0$, and $B_{\phi,\mathrm{G}}=3.2 \times 10^{-7} B_0$,
where $B_0$ is defined in \sref{model:units}. These values are
comparable to the magnitude of the polar magnetic field
$B_\mathrm{p}=2.9 \times 10^{-7} B_0$.

\subsection{Dependence on curvature}
\label{sec:stab3d:real}
As discussed in \sref{model:units},
PM07 argued that reducing $R_\ast$ and $M_\ast$ does
not affect the equilibrium structure as long as $h_0$ remains
constant, at least in the small-$M_a$ limit. To test whether this
also holds for the dynamical behaviour of the system, we perform two
runs, models J and K, with $a=75$ and $a=100$ respectively.

\begin{figure}
  \includegraphics[width=84mm, keepaspectratio]{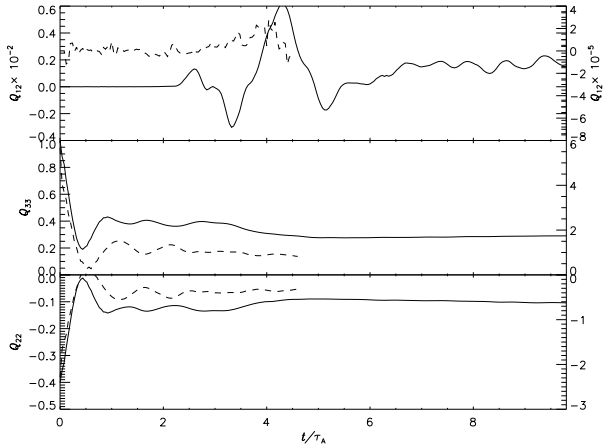}
  \caption{Mass quadrupole moments $Q_{ij}$ for model J ($a=75$, solid
    curve) and model K ($a=100$, dashed curve), plotted as a function
    of time (in units of the Alfv\'{e}n time). The scale for model J (K)
    appears on the left (right) vertical axis. Although the
    transition phase  is less distinct than in
    \fref{stab:modd_quadru}, these models basically share the same
    dynamics as lower curvature runs.}
  \label{fig:stab:modelcurvature_quadrupole}
\end{figure}

\fref{stab:modelcurvature_quadrupole} plots $Q_{ij}$ versus time for these
models. The transition phase is more gradual than model D
(\fref{stab:modd_quadru}). Again, however,  $Q_{12}$ rises
significantly, marking a deviation from axisymmetry. \citet{Melatos05}
found $\epsilon \propto a^2$ analytically in the small-$M_a$ regime,
so we fit a parabola to the simulation data (for $M_a=M_c$):
\begin{equation}
  \label{eq:stab3d:curve_scaling}
  \frac{\epsilon}{10^{-13}} = 1.82 a^2.
\end{equation}

A realistic star has $a=1.9 \times 10^4$
(cf. \sref{model:units}). Extrapolating (\ref{eq:stab3d:curve_scaling}),
we find $\epsilon=6.6 \times 10^{-5}$. (An ellipticity this large is
close to the upper limit inferred from existing gravitational-wave
nondetections; see \sref{discussion} for more details.) However, it
should be remembered that \eref{stab3d:curve_scaling} is an
overestimate, because the computations in this paper neglect nonideal
MHD effects.

\section{Global MHD oscillations}
\label{sec:oscillations}
In this section, we explore the natural oscillation modes of a
nonaxisymmetric magnetic mountain. We do this by loading the final
state from models D, F, and G into \textsc{zeus-mp} and setting
$\mathbf{v}=0$ on the whole grid. This procedure introduces numerical
perturbations that are sufficient to excite small linear oscillation
modes, albeit an uncontrolled distribution thereof. We then compute
the power spectrum
\begin{equation}
  P[S] (\omega) = \left|\frac{1}{N} \sum_{i=0}^{N-1} S(t_i)
    \mathrm{e}^{-i \omega t_i/N}\right|^2 
\end{equation}
by evaluating the discrete Fourier transform of the scalar
function $S(t)$ [e.g. $B_r(t)$] at $N$ sample times $t_i$.

In order to explore the magnetic modes, we examine $B_r$, $B_\theta$, and $B_\phi$.
We choose one point on the grid where the amplitude of the oscillations is
high, namely $(r, \theta, \phi)=(50.03 h_0, 0.26, 2.06)$ and compute
$P[B_r]$, $P[B_\theta]$, and $P[B_\phi]$. The results are displayed in
\fref{osc:mass10}. We can distinguish five different spectral peaks  at
$10^{3} \omega \tau_0=0.78, 1.1, 1.2, 2.1, 3.2$, which are
more or less distinct for the different components.

\begin{figure}
  \includegraphics[width=84mm,keepaspectratio]{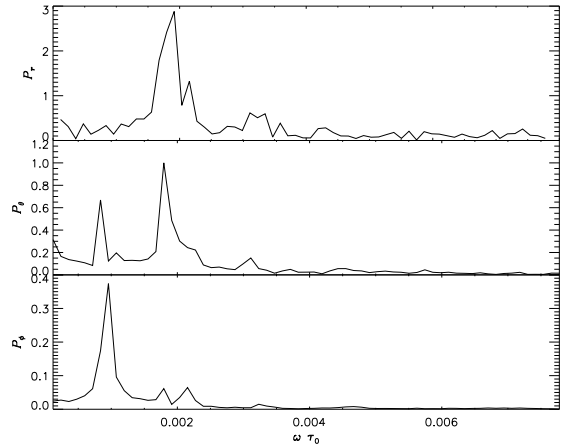}
  \caption{Power spectrum of $B_r$, $B_\theta$, and $B_\phi$ (top to
    bottom) for model D (arbitrary units), plotted as a function of
    angular Fourier frequency (in units of $\tau_0^{-1}$).}
  \label{fig:osc:mass10}
\end{figure}

For a magnetized gravitating slab in a plane-parallel geometry, one can
distinguish three different MHD modes \citep{Goedbloed04}: slow
magnetosonic, Alfv\'{e}n, and fast magnetosonic. Each mode consists of
a discrete set of eigenmodes and a continuous spectrum, which
are clearly separated. Unfortunately, such clean separation cannot be
expected for a highly inhomogenous plasma in spherical geometry. Generally,
different parts of the spectrum overlap or degenerate into a single
point in a nontrivial way. We therefore restrict the discussion below to
some qualitative remarks.

The MHD spectrum contains genuine singularities, when the
eigenfrequency coincides with the Alfv\'{e}n or slow magneto-sonic
frequency at some location within the magnetic mountain. In this 
case, the boundary value problem becomes singular; the boundary
conditions can be fulfilled for a continuous range of
frequencies. The singular frequencies depend on the components of the
wave vector perpendicular to the direction of inhomogenity.

It is unclear whether the band $\omega<0.002 \tau_0^{-1}$, which looks
``filled'' in \fref{osc:mass10}, belongs to the
continuous part of the spectrum or else is an artifact of the nonzero line
width from numerical damping (which can be estimated from the
sample times $t_i$ to be $\sim 1.2 \times 10^{-4} \omega \tau_0$). We
do not observe any singular behaviour in the field variables, but we
note that singularities would be suppressed by the shock-capturing
algorithm (i.e. the artificial viscosity) in \textsc{zeus-mp}. We
conclude that the features in \fref{osc:mass10} are probably discrete
lines.

\begin{figure}
  \includegraphics[width=84mm,keepaspectratio]{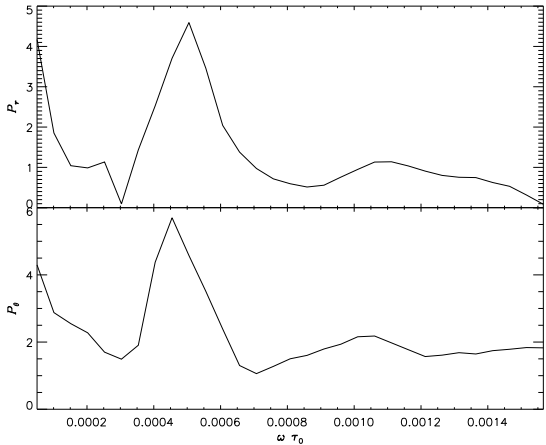}
  \caption{Power spectra of $B_r$ and $B_\theta$ (top to
    bottom) for model A (arbitrary units), plotted as a function of
    angular Fourier frequency (in units of $\tau_0^{-1}$).}
  \label{fig:osc:mass10_25d}
\end{figure}

Let us compare these results to the spectrum of the axisymmetric model A
(\fref{osc:mass10_25d}). We first note that the
Alfv\'{e}n frequency $\omega_A=0.018 \tau_0^{-1} (M_a/M_c)^{-1/2}$ and
acoustic frequency $\omega_s=0.48 \tau_0^{-1}$ found by PM07 are outside the
range of this plot, which is set by the Nyquist frequency
$\omega_\mathrm{N}=(4 \pi N \Delta t)^{-1}$ ($\Delta t=50 \tau_0$ for
model A and $\Delta t=10 \tau_0$ for models D--G). Here, we
are restricted to low frequency oscillations which are generally
associated with global magnetic modes. Most distinct is the peak at
$10^3 \omega \tau_0=0.5$, which is not visible in
\fref{osc:mass10}. This long wavelength poloidal mode is suppressed in
favor of toroidal modes in the three-dimensional
configuration. However, the small peak at $10^3 \omega \tau_0=1.1$ is
present in both systems. This example illustrates vividly how relaxing the
axisymmetric constraint leads to a different MHD spectrum.

\begin{figure}
  \includegraphics[width=84mm,keepaspectratio]{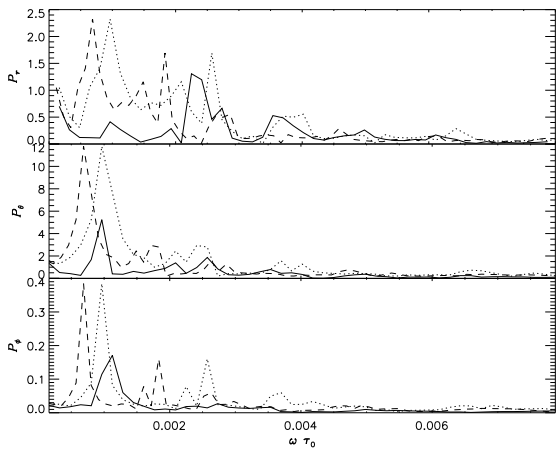}
  \caption{Power spectra of $B_r$, $B_\theta$, and $B_\phi$ (top to
    bottom) for models F (solid curve) and G (dashed curve) in
    arbitrary units, plotted as a function of
    angular Fourier frequency in units of $\tau_0^{-1}$. We overplot
    the spectrum of model G, stretched by a factor of 1.4 in $\omega$,
    as dotted curve, by way of comparison.}
  \label{fig:osc:massvar}
\end{figure}

\fref{osc:massvar} shows the power spectrum for models F (solid) and G
(dashed). The most distinct peaks are again concentrated in the low
frequency region. We can roughly match the peaks of models G
and F by stretching the former spectrum by a factor of 1.4 in
frequency. The higher $M_a$ equilibrium has a similar structure, but
the Alfv\'{e}n timescale is lower because the plasma density is 85 per cent
higher.

A complete analytic determination of the discrete and continuous
components of the MHD spectrum via a full linear mode analysis will be
attemped in a forthcoming paper.

\section{Discussion}
\label{sec:discussion}
Magnetically confined mountains on accreting neutron stars screen
the magnetic dipole moment of the star. Potentially, therefore, the
process of polar magnetic burial can explain the observed reduction of
$\mu$ with $M_a$ in neutron stars with an accretion history. However,
before magnetic burial can be invoked as a viable explanation,
the question of stability must be resolved. In this article, we
concentrate on the important aspect of three-dimensional stability,
deferring resistive processes to future work (especially the issue of
resistive g-modes\footnote{J. Arons, private communication}).

We find that the axisymmetric configurations in PM04 are susceptible
to the three-dimensional magnetic buoyancy instability. The
instability proceeds via the undular submode, with growth rate
$\propto \lambda^{1/2}$, limited by the toroidal grid
resolution. However, instead of breaking up and 
reverting to an isothermal atmosphere threaded by a dipolar magnetic
field, the magnetic field reconfigures (over a few Alfv\'{e}n times) and
settles down into a new nonaxisymmetric equilibrium which is still
highly distorted. Just as the axisymmetric solutions in PM04 are the final 
saturated states of the nonlinear evolution of the Parker instability
in two dimensions, we find here the three-dimensional
equivalent. This surprising result is the main conclusion of the
paper. It holds irrespective of the outer boundary condition and
curvature rescaling factor, but it depends critically on the
line-tying boundary condition at the stellar surface.

The final state is predominantly axisymmetric, with $1.5 \le
|Q_{12}/Q_{33}|/10^{-3} \le 3.2$ for models D, F, and G ($0.6 \le
M_a/M_c \le 1.4$). The ellipticity for model G reaches $1.6 \times
10^{-10}$ in the downscaled star. 

\begin{figure}
  \includegraphics[width=84mm,keepaspectratio]{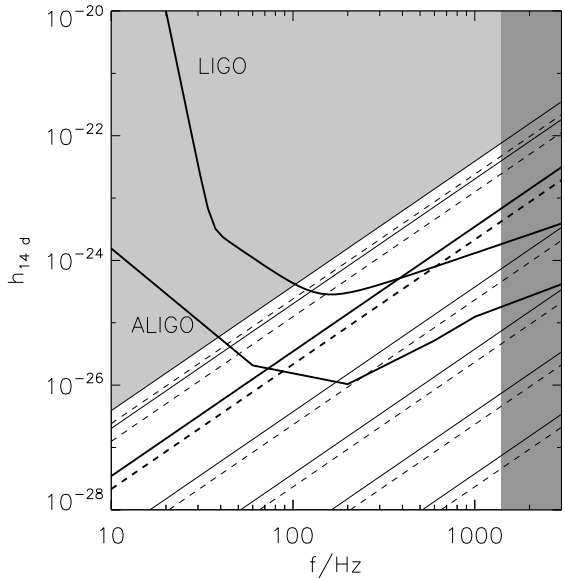}
  \caption{Amplitude of the gravitational wave signal $h_0$ for $M_a/M_\odot= 10^{-9},
     10^{-8},10^{-7},10^{-6},10^{-5},10^{-4},10^{-3}$, for the
     axisymmetric (solid lines) and nonaxisymmetric
     equilibrium (dashed lines). The sensitivities of Initial and
     Advanced LIGO, assuming 14 days coherent integration, are also
     plotted (upper and lower curves respectively).
     The growth of the mountain is arrested for $M_a
     \ga 1.2 \times 10^{-5} M_\odot$ (light shaded region), due to
     Ohmic dissipation \citep{Melatos05, Vigelius08}, while the right-hand edge is excluded at
     present because no accreting millisecond pulsars have been
     discovered with $f_\ast > 0.7$ kHz (dark shaded region).}
  \label{fig:dis:detectability}
\end{figure}

The stability of magnetic mountains is important for the emission of
gravitational waves from accreting millisecond pulsars, as pointed out
previously by \citet{Melatos05}. Persistent X-ray pulsations from accreting binary
pulsars imply that the angle between the spin vector
$\mathbf\Omega$ and the magnetic symmetry axis $\bmu$ is not
zero \citep{Romanova04, Kulkarni05}. Hence a magnetic mountain constitutes a time-varying mass
quadrupole which emits gravitational waves. Furthermore, the star
precesses in general, emitting gravitational waves at the spin
frequency and its first harmonic. The amplitude of the resulting signal (with curvature
upscaled to a realistic neutron star at a distance $d=10$ kpc using
$\epsilon \propto a^2$) is plotted in \fref{dis:detectability} for
$10^{-9} \le M_a/M_\odot \le 10^{-3}$. The amplitude of the average
signal that can be detected by the Laser Interferometer Gravitational
Wave Observatory (LIGO) from a periodic source with a false alarm rate of
1 per cent and a false dismissal rate of 10 per cent over an
integration time of $T_0=14$ days \citep{JKSI, Abbott2004}, is
overplotted in \fref{dis:detectability}. This $T_0$ can realistically
be achieved computationally.

At this point, it is important to acknowledge that $M_a=1.4
M_c \approx 1.7 \times 10^{-4} M_\odot$ is still well below $M_a \sim
0.1 M_\odot$, the mass required to spin up a neutron star to millisecond periods
\citep{Burderi99}. At present, this high-mass regime is
not accessible numerically; neither the GS solver nor
\textsc{zeus-mp} can handle the steep magnetic gradients involved.
By the same token, Ohmic diffusion becomes important in
this high-$M_a$ regime \citep{Melatos05, Vigelius08}, smoothing the gradients
and mitigating the numerical challenge. We postpone studying
realistic values of $M_a$ to an accompanying paper, which will
concentrate on non-ideal MHD simulations. However, to make a rough
estimate regarding detectability here, we assume that non-ideal
effects stall the growth of the mountain at $M_a \approx
M_c$, following \citet{Melatos05}. This includes the region shaded light grey in
\fref{dis:detectability}. Furthermore, no accreting millisecond
pulsars have been discovered spinning faster than $f_\ast \ga 720$ Hz,
possibly due to braking by gravitational waves
\citep{Bildsten98, Chakrabarty03}. The region with $2 f_\ast \ga 1400$ Hz is shaded
dark grey in \fref{dis:detectability}.

Even with those exclusions, \fref{dis:detectability} demonstrates
that there is a fair prospect of detecting gravitational waves from
accreting X-ray millisecond pulsars in the near future, for accreted
masses as low as $M_a \approx 10^{-4} M_\odot$. Recent directed
searches for gravitational waves from the nearby X-ray source Sco-X1
found no signal at the level $h_0 \ga 10^{-22}$ \citep{LIGO06}, thereby setting an upper bound on
the ellipticity of $\epsilon=3.6 \times 10^{-3}$.

\bibliography{stability.bib}

\begin{thebibliography}{}

\bibitem[\protect\citeauthoryear{{{Abbott}, B. {et~al.}}}{{{Abbott}, B.
  {et~al.}}}{2004}]{Abbott2004}
{{Abbott}, B. {et~al.}} 2004, \prd, 69, 082004

\bibitem[\protect\citeauthoryear{{{Abbott}, B. {et~al.}}}{{{Abbott}, B.
  {et~al.}}}{2007}]{LIGO06}
{{Abbott}, B. {et~al.}} 2007, \prd, 76, 082001

\bibitem[\protect\citeauthoryear{{Akiyama}, {Wheeler}, {Meier} \&
  {Lichtenstadt}}{{Akiyama} et~al.}{2003}]{Akiyama03}
{Akiyama} S.,  {Wheeler} J.~C.,  {Meier} D.~L.,    {Lichtenstadt} I.,  2003,
  \apj, 584, 954

\bibitem[\protect\citeauthoryear{{Balbus} \& {Hawley}}{{Balbus} \&
  {Hawley}}{1998}]{Balbus98}
{Balbus} S.~A.,  {Hawley} J.~F.,  1998, Rev. Mod. Phys., 70, 1

\bibitem[\protect\citeauthoryear{{Bildsten}}{{Bildsten}}{1998}]{Bildsten98}
{Bildsten} L.,  1998, \apjl, 501, L89+

\bibitem[\protect\citeauthoryear{{Biskamp}}{{Biskamp}}{1993}]{Biskamp93}
{Biskamp} D.,  1993, {Nonlinear magnetohydrodynamics}.
Cambridge University Press, Cambridge.

\bibitem[\protect\citeauthoryear{{Bisnovatyi-Kogan} \&
  {Komberg}}{{Bisnovatyi-Kogan} \& {Komberg}}{1974}]{Bisnovatyi74}
{Bisnovatyi-Kogan} G.~S.,  {Komberg} B.~V.,  1974, Soviet Astronomy, 18, 217

\bibitem[\protect\citeauthoryear{{Bonazzola} \& {Gourgoulhon}}{{Bonazzola} \&
  {Gourgoulhon}}{1996}]{Bonazzola96}
{Bonazzola} S.,  {Gourgoulhon} E.,  1996, \aap, 312, 675

\bibitem[\protect\citeauthoryear{{Bouwkamp} \& {Casimir}}{{Bouwkamp} \&
  {Casimir}}{1954}]{Bouwkamp54}
{Bouwkamp} C.~J.,  {Casimir} H.~B.~G.,  1954, Physica, 20, 539

\bibitem[\protect\citeauthoryear{{Brown} \& {Bildsten}}{{Brown} \&
  {Bildsten}}{1998}]{Brown98}
{Brown} E.~F.,  {Bildsten} L.,  1998, \apj, 496, 915

\bibitem[\protect\citeauthoryear{{Burderi}, {Possenti}, {Colpi}, {di Salvo} \&
  {D'Amico}}{{Burderi} et~al.}{1999}]{Burderi99}
{Burderi} L.,  {Possenti} A.,  {Colpi} M.,  {di Salvo} T.,    {D'Amico} N.,
  1999, \apj, 519, 285

\bibitem[\protect\citeauthoryear{{Chakrabarty}, {Morgan}, {Muno}, {Galloway},
  {Wijnands}, {van der Klis} \& {Markwardt}}{{Chakrabarty}
  et~al.}{2003}]{Chakrabarty03}
{Chakrabarty} D.,  {Morgan} E.~H.,  {Muno} M.~P.,  {Galloway} D.~K.,
  {Wijnands} R.,  {van der Klis} M.,    {Markwardt} C.~B.,  2003, \nat, 424, 42

\bibitem[\protect\citeauthoryear{{Cumming}, {Arras} \& {Zweibel}}{{Cumming}
  et~al.}{2004}]{Cumming04}
{Cumming} A.,  {Arras} P.,    {Zweibel} E.,  2004, \apj, 609, 999

\bibitem[\protect\citeauthoryear{{Cutler}}{{Cutler}}{2002}]{Cutler2002}
{Cutler} C.,  2002, \prd, 66, 084025

\bibitem[\protect\citeauthoryear{{Geppert} \& {Rheinhardt}}{{Geppert} \&
  {Rheinhardt}}{2002}]{Geppert02}
{Geppert} U.,  {Rheinhardt} M.,  2002, \aap, 392, 1015

\bibitem[\protect\citeauthoryear{{Geppert} \& {Urpin}}{{Geppert} \&
  {Urpin}}{1994}]{Geppert94}
{Geppert} U.,  {Urpin} V.,  1994, \mnras, 271, 490

\bibitem[\protect\citeauthoryear{{Goedbloed} \& {Halberstadt}}{{Goedbloed} \&
  {Halberstadt}}{1994}]{Goedbloed94}
{Goedbloed} J.~P.,  {Halberstadt} G.,  1994, \aap, 286, 275

\bibitem[\protect\citeauthoryear{{Goedbloed} \& {Poedts}}{{Goedbloed} \&
  {Poedts}}{2004}]{Goedbloed04}
{Goedbloed} J.~P.~H.,  {Poedts} S.,  2004, {Principles of
  Magnetohydrodynamics}.
Cambridge University Press, Cambridge.

\bibitem[\protect\citeauthoryear{{Greene} \& {Johnson}}{{Greene} \&
  {Johnson}}{1968}]{Greene68}
{Greene} J.~M.,  {Johnson} J.~L.,  1968, Plasma Physics, 10, 729

\bibitem[\protect\citeauthoryear{{Haskell}, {Jones} \& {Andersson}}{{Haskell}
  et~al.}{2006}]{Haskell06}
{Haskell} B.,  {Jones} D.~I.,    {Andersson} N.,  2006, \mnras, 373, 1423

\bibitem[\protect\citeauthoryear{{Hawley} \& {Stone}}{{Hawley} \&
  {Stone}}{1995}]{Hawley95}
{Hawley} J.~F.,  {Stone} J.~M.,  1995, Comp. Phys. Comm., 89, 127

\bibitem[\protect\citeauthoryear{{Hayes}, {Norman}, {Fiedler}, {Bordner}, {Li},
  {Clark}, {ud-Doula} \& {Mac Low}}{{Hayes} et~al.}{2006}]{Hayes06}
{Hayes} J.~C.,  {Norman} M.~L.,  {Fiedler} R.~A.,  {Bordner} J.~O.,  {Li}
  P.~S.,  {Clark} S.~E.,  {ud-Doula} A.,    {Mac Low} M.-M.,  2006, \apjs, 165,
  188

\bibitem[\protect\citeauthoryear{{Hughes} \& {Cattaneo}}{{Hughes} \&
  {Cattaneo}}{1987}]{Hughes87}
{Hughes} D.~W.,  {Cattaneo} F.,  1987, Geophysical and Astrophysical Fluid
  Dynamics, 39, 65

\bibitem[\protect\citeauthoryear{{Jackson}}{{Jackson}}{1998}]{Jackson98}
{Jackson} J.~D.,  1998, {Classical Electrodynamics}.
Wiley-VCH, New York.

\bibitem[\protect\citeauthoryear{{Jaranowski}, {Kr{\'o}lak} \&
  {Schutz}}{{Jaranowski} et~al.}{1998}]{JKSI}
{Jaranowski} P.,  {Kr{\'o}lak} A.,    {Schutz} B.~F.,  1998, \prd, 58, 063001

\bibitem[\protect\citeauthoryear{{Konar} \& {Bhattacharya}}{{Konar} \&
  {Bhattacharya}}{1997}]{Konar97}
{Konar} S.,  {Bhattacharya} D.,  1997, \mnras, 284, 311

\bibitem[\protect\citeauthoryear{{Kulkarni} \& {Romanova}}{{Kulkarni} \&
  {Romanova}}{2005}]{Kulkarni05}
{Kulkarni} A.~K.,  {Romanova} M.~M.,  2005, \apj, 633, 349

\bibitem[\protect\citeauthoryear{{Lifschitz}}{{Lifschitz}}{1989}]{Lifschitz89}
{Lifschitz} A.~E.,  1989, {Magnetohydrodynamics and Spectral Theory}.
Kluwer Academic Publishers, London.

\bibitem[\protect\citeauthoryear{{Litwin}, {Brown} \& {Rosner}}{{Litwin}
  et~al.}{2001}]{Litwin01}
{Litwin} C.,  {Brown} E.~F.,    {Rosner} R.,  2001, \apj, 553, 788

\bibitem[\protect\citeauthoryear{{Lovelace}, {Romanova} \&
  {Bisnovatyi-Kogan}}{{Lovelace} et~al.}{2005}]{Lovelace2005}
{Lovelace} R.~V.~E.,  {Romanova} M.~M.,    {Bisnovatyi-Kogan} G.~S.,  2005,
  \apj, 625, 957

\bibitem[\protect\citeauthoryear{{Masada}, {Sano} \& {Takabe}}{{Masada}
  et~al.}{2006}]{Masada06}
{Masada} Y.,  {Sano} T.,    {Takabe} H.,  2006, \apj, 641, 447

\bibitem[\protect\citeauthoryear{{Matsumoto} \& {Shibata}}{{Matsumoto} \&
  {Shibata}}{1992}]{Matsumoto92}
{Matsumoto} R.,  {Shibata} K.,  1992, \pasj, 44, 167

\bibitem[\protect\citeauthoryear{{Melatos} \& {Payne}}{{Melatos} \&
  {Payne}}{2005}]{Melatos05}
{Melatos} A.,  {Payne} D.~J.~B.,  2005, \apj, 623, 1044

\bibitem[\protect\citeauthoryear{{Melatos} \& {Phinney}}{{Melatos} \&
  {Phinney}}{2001}]{Melatos01}
{Melatos} A.,  {Phinney} E.~S.,  2001, Publ. Astronom. Soc. Aust., 18, 421

\bibitem[\protect\citeauthoryear{{Mouschovias}}{{Mouschovias}}{1974}]{Mouschov%
ias74}
{Mouschovias} T.~C.,  1974, \apj, 192, 37

\bibitem[\protect\citeauthoryear{{Muslimov} \& {Tsygan}}{{Muslimov} \&
  {Tsygan}}{1985}]{Muslimov85}
{Muslimov} A.~G.,  {Tsygan} A.~I.,  1985, Sov. Astron. Lett., 11, 80

\bibitem[\protect\citeauthoryear{{Owen}}{{Owen}}{2006}]{Owen06}
{Owen} B.~J.,  2006, Classical and Quantum Gravity, 23, 1

\bibitem[\protect\citeauthoryear{{Payne}}{{Payne}}{2005}]{Payne05}
{Payne} D.~J.~B.,  2005, PhD thesis, School of Physics. University of
  Melbourne.

\bibitem[\protect\citeauthoryear{{Payne} \& {Melatos}}{{Payne} \&
  {Melatos}}{2004}]{Payne04}
{Payne} D.~J.~B.,  {Melatos} A.,  2004, \mnras, 351, 569

\bibitem[\protect\citeauthoryear{{Payne} \& {Melatos}}{{Payne} \&
  {Melatos}}{2006a}]{Payne06a}
{Payne} D.~J.~B.,  {Melatos} A.,  2006a, \apj, 641, 471

\bibitem[\protect\citeauthoryear{{Payne} \& {Melatos}}{{Payne} \&
  {Melatos}}{2006b}]{Payne07c}
{Payne} D.~J.~B.,  {Melatos} A.,  2006b, \apj, 652, 597

\bibitem[\protect\citeauthoryear{{Payne} \& {Melatos}}{{Payne} \&
  {Melatos}}{2007}]{Payne06b}
{Payne} D.~J.~B.,  {Melatos} A.,  2007, \mnras, 376, 609

\bibitem[\protect\citeauthoryear{{Pons} \& {Geppert}}{{Pons} \&
  {Geppert}}{2007}]{Pons07}
{Pons} J.~A.,  {Geppert} U.,  2007, \aap, 470, 303

\bibitem[\protect\citeauthoryear{{Priest}}{{Priest}}{1984}]{Priest84}
{Priest} E.~R.,  1984, {Solar magneto-hydrodynamics}.
Geophysics and Astrophysics Monographs, Dordrecht: Reidel.

\bibitem[\protect\citeauthoryear{{Romani}}{{Romani}}{1990}]{Romani90}
{Romani} R.~W.,  1990, \nat, 347, 741

\bibitem[\protect\citeauthoryear{{Romanova}, {Ustyugova}, {Koldoba} \&
  {Lovelace}}{{Romanova} et~al.}{2004}]{Romanova04}
{Romanova} M.~M.,  {Ustyugova} G.~V.,  {Koldoba} A.~V.,    {Lovelace} R.~V.~E.,
   2004, \apj, 610, 920

\bibitem[\protect\citeauthoryear{{Schindler}, {Hesse} \& {Birn}}{{Schindler}
  et~al.}{1988}]{Schindler88}
{Schindler} K.,  {Hesse} M.,    {Birn} J.,  1988, \jgr, 93, 5547

\bibitem[\protect\citeauthoryear{{Shapiro} \& {Teukolsky}}{{Shapiro} \&
  {Teukolsky}}{1983}]{Shapiro83}
{Shapiro} S.~L.,  {Teukolsky} S.~A.,  1983, {Black holes, white dwarfs, and
  neutron stars: The physics of compact objects}.
Wiley-Interscience, New York.

\bibitem[\protect\citeauthoryear{{Srinivasan}, {Bhattacharya}, {Muslimov} \&
  {Tsygan}}{{Srinivasan} et~al.}{1990}]{Srinivasan90}
{Srinivasan} G.,  {Bhattacharya} D.,  {Muslimov} A.~G.,    {Tsygan} A.~J.,
  1990, Curr. Sci., 59, 31

\bibitem[\protect\citeauthoryear{{Taam} \& {van de Heuvel}}{{Taam} \& {van de
  Heuvel}}{1986}]{Taam86}
{Taam} R.~E.,  {van de Heuvel} E.~P.~J.,  1986, \apj, 305, 235

\bibitem[\protect\citeauthoryear{{Urpin} \& {Konenkov}}{{Urpin} \&
  {Konenkov}}{1997}]{Urpin97}
{Urpin} V.,  {Konenkov} D.,  1997, \mnras, 284, 741

\bibitem[\protect\citeauthoryear{{van den Heuvel} \& {Bitzaraki}}{{van den
  Heuvel} \& {Bitzaraki}}{1995}]{vanDenHeuvel95}
{van den Heuvel} E.~P.~J.,  {Bitzaraki} O.,  1995, \aap, 297, L41+

\bibitem[\protect\citeauthoryear{{Vigelius} \& {Melatos}}{{Vigelius} \&
  {Melatos}}{2008}]{Vigelius08}
{Vigelius} M.,  {Melatos} A.,  2008, in preparation

\bibitem[\protect\citeauthoryear{{Wijers}}{{Wijers}}{1997}]{Wijers97}
{Wijers} R.~A.~M.~J.,  1997, \mnras, 287, 607

\bibitem[\protect\citeauthoryear{{Zhang}}{{Zhang}}{1998}]{Zhang98}
{Zhang} C.~M.,  1998, \apss, 262, 97

\end{thebibliography}

\appendix
\section{Defining the grid and boundary conditions in
  \textsc{zeus-mp}}
\label{sec:app:zeusvars}

In this appendix, we briefly outline the key variables and settings in
\textsc{zeus-mp}, to aid the reader in reproducing our numerical results.
Our grid consists of \texttt{ggen1:nbxl}, \texttt{ggen2:nbxl}, and
\texttt{ggen3:nbxl} blocks in the $r$, $\theta$, and $\phi$ direction,
respectively. The integration volume is defined by $R_\ast/h_0 \leq r
\leq R_m$, $0 \leq \theta \leq \pi/2$, $0 \leq \phi < 2\pi$. The
radial coordinate in the GS code, $\tilde{x}=(r-R_\ast)/h_0$ ($0 \leq
\tilde{x} \leq X$), is stretched logarithmically according to
$\tilde{x}_1=\log(\tilde{x}+\mathrm{e}^{-L_x})+L_x$, where
$L_x$ controls the zooming (PM04). This grid is implemented
by setting the \textsc{zeus-mp} parameters $\mathtt{ggen1:x1min}=R_\ast/h_0$ and
$\mathtt{ggen1:x1max}=\mathtt{ggen1:x1min}+X$. Stretching is achieved
via the parameter \texttt{ggen1:x1rat}, which sets the radial length ratio
of two neighbouring zones. In order to get consistent radial grid
positions in the GS code and \textsc{zeus-mp}, we set
$\mathtt{ggen1:x1rat}=(X \mathrm{e}^{L_x}+1)^{(G_x-1)^{-1}}$.

Boundary conditions are enforced in \textsc{zeus-mp} via ghost cells,
which frame the active grid cells. Several predefined prescriptions
are supplied to implement a variety of standard boundary conditions. In the
$\phi$ direction, we choose periodic boundary conditions [\texttt{ikb.niks(1)=4} and
\texttt{okb.noks(1)=4}]. The $\theta=\pi/2$ surface is reflecting,
with normal magnetic field [\texttt{ojb.nojs(1)= 5}], which translates to
$\mathbf{v}_\perp=\mathbf{B}_\parallel=0$. The line $\theta=0$
is also reflecting [\texttt{ijb.nijs(1)= -1}] with tangential magnetic field
($\mathbf{v}_\perp=\mathbf{B}_\perp=0$). Additionally, the toroidal
component $B_\phi$ is reversed at the boundary,
i.e. $B^<_\phi=-B^>_\phi$, where $B_\phi^<$ and $B_\phi^>$ are the
field components for $\theta<0$ and $\theta>0$, respectively. The
outer surface $r=R_m$ is usually an outflow [\texttt{oib
  nois(1)= 2}] boundary, i.e. zero gradient. The stellar surface
is impenetrable, so the inner $r=R_\ast$ boundary is inflow
[\texttt{iib.niis(1)= 3}]. This enables us to impose line-tying at
$r=R_\ast$ by fixing the density and magnetic field there. We also use
an isothermal equation of state (\texttt{XISO=.true.}).

\section{Mass multipole moments}
\label{sec:app:multipole}
We work out the mass quadrupole moment in Cartesian coordinates from
the code output in spherical coordinates. Following \citet{Jackson98},
we define the spherical mass multipole moments according to
\begin{equation}
  \label{eq:multipole:sph_components}
  q_{lm}=\int \mathrm{d}^3 \mathbf{x'} Y^*_{lm}(\theta',\phi') r'^l \rho(\mathbf{x}'),
\end{equation}
where $Y_{lm}$ denotes the usual orthonormal set of spherical
harmonics.

The spherical quadrupole moments are related to the traceless,
Cartesian quadrupole moment tensor,
\begin{equation}
  Q_{ij}=\int d^3 x' \, (3 x_i' x_j'-r'^2 \delta_{ij}) \rho(\mathbf{x'}),
\end{equation}
by  $Q_{11}=6 (2 \pi/15)^{1/2} \mathrm{Re}(q_{22})-2 (4 \pi/5)^{1/2} q_{20}$,  $Q_{12}=-6 (2 \pi/15)^{1/2}
\mathrm{Im}(q_{22})$, $Q_{13}=-3 (8 \pi/15)^{1/2}
\mathrm{Re}(q_{21})$, $Q_{22}=-6 (2\pi/15)^{1/2} \mathrm{Re}(q_{22}) -
(4\pi/5)^{1/2} q_{20}$, and $Q_{23}=3 (8\pi/15)^{1/2} \mathrm{Im}(q_{21})$.

In the axisymmetric case (when the star and the mountain form a
prolate spheroid), we have $\rho=\rho(r, \theta)$ and the
$\phi$ integrals in \eeref{multipole:sph_components} vanish. $Q$ is
then diagonal with components
$Q_{\hat{x}\hat{x}}=Q_{\hat{y}\hat{y}}=-Q_{\hat{z}\hat{z}}/2$, with
respect to the body coordinate system, and we
can introduce the ellipticity\footnote{Note that the
  definition of the ellipticity varies in the literature. The
  ellipticity defined here is consistent with \citet{Bonazzola96} and
  \citet{Shapiro83} and is related to the ellipticity in
  \citet{Melatos05} and \citet{JKSI} by $|\epsilon|=3
  \epsilon_\mathrm{MP}$. \citet{LIGO06} used a different ellipticity
  defined for a \emph{triaxial} rotator, $\epsilon=(I_{xx}-I_{yy})/I_{zz}$.} $\epsilon$, where we assume
that the $\hat{z}$ axis is the symmetry axis:
\begin{equation}
  \label{eq:app:gw:ellipticity_gen}
  \epsilon=   \frac{Q_{\hat{z}\hat{z}}}{2 I_{\hat{z}\hat{z}}} =
  \frac{3 (I_{\hat{z}\hat{z}}-I_{\hat{x}\hat{x}})}{I_{\hat{z}\hat{z}}},
\end{equation}
with $I_{\hat{z}\hat{z}}=2 M_\ast R_\ast^2/5$ is the moment of inertia
along the rotation axis for a biaxial ellipsoid with mass $M_\ast$ and
minor axis $R_\ast$.
We can compute the ellipticity directly from the code output through
\begin{equation}
  \label{eq:app:gw:ellipticity}
  \epsilon=\frac{1}{2 I_{\hat{z}\hat{z}}}\int d\theta\,d\phi\,dr\, \rho r^4 \sin \theta  (3 \cos^2
  \theta - 1).
\end{equation}

\section{Magnetic multipole moments}
\label{sec:app:magmulti}
In a source-free region $\mathbf{J}=0$, a magnetic field $\mathbf{B}$
is determined solely by its radial component $B_r$
\citep{Bouwkamp54}, which, from Maxwell's equations, satisfies
the Laplace equation
\begin{equation}
  \label{eq:app:laplace}
  \nabla^2 B_r=0.
\end{equation}
One can therefore define the magnetic multipoles as the expansion coefficients
in the general solution of the boundary value problem
\eeref{app:laplace}, viz.
\begin{equation}
  B_r=\sum_{l=0}^{\infty} \sum_{m=-l}^l
  d_{lm} r^{-(l+1)} Y_{lm} (\theta, \phi),
\end{equation}
with
\begin{equation}
  d_{lm} = r^{l+1} \int \mathrm{d}\Omega\; Y_{lm}^\ast
   \mathbf{r}\bcdot\mathbf{B}.
\end{equation}

Note that $d_{10}$ is related to the magnetic moment $\mu$ of a dipole
field $\mathbf{B}(\mathbf{r})=\mu r^{-3} (2 \cos \theta \mathbf{e}_r +
\sin \theta \mathbf{e}_\theta)$ by $d_{10}=4 (\pi/3)^{1/2} \mu$.

In the case of north-south symmetry, we find $d_{10} = 2 \hat{d}_{10}$
and $d_{21}=2 \hat{d}_{21}$, where a hat denotes the moment evaluated
on the hemisphere. All other coefficients vanish.
\end{document}